\documentclass[a4paper,12pt]{article}
\usepackage{graphicx,amssymb,bm,latexsym,color,epsf}
\usepackage{cite}
\usepackage{amsmath}
\usepackage{appendix}
\usepackage[all]{xy}

\makeatletter
\@addtoreset{equation}{section}

\makeatother
\newcommand{\bea}{\begin{eqnarray}}
\newcommand{\eea}{\end{eqnarray}}
\newcommand{\vs}[1]{\vspace{#1 mm}}
\newcommand{\hs}[1]{\hspace{#1 mm}}
\renewcommand{\a}{\alpha}
\renewcommand{\b}{\beta}

\renewcommand{\d}{\delta}
\newcommand{\e}{\epsilon}
\newcommand{\s}{\sigma}
\renewcommand{\t}{\theta}

\newcommand{\la}{\lambda}
\newcommand{\pa}{\partial}
\newcommand{\nn}{\nonumber\\}
\newcommand{\p}[1]{(\ref{#1})}
\newcommand{\lan}{\langle}
\newcommand{\ran}{\rangle}

\newcommand{\VEV}[1]{\langle0| #1 |0\rangle}
\newcommand{\EV}[1]{\langle#1 \rangle}
\newcommand{\B}{{\rm B}}

\newcommand{\FP}{{\rm FP}}
\newcommand{\T}{{\rm T}}

\newcommand{\bfp}{{\boldsymbol p}}

\newcommand{\calB}{{\cal B}\kern-.70em{\cal B}}
\newcommand{\calC}{{\cal C}\kern-.58em{\cal C}}
\newcommand{\calD}{{\cal D}}
\newcommand{\half}{(\kern-.3pt1\kern-.8pt/\kern-.2pt2)}

\newcommand{\calT}{{\cal T}}
\newcommand{\mbf}[1]{{\boldsymbol #1}}
\newcommand{\rtg}{{\sqrt{-g}\,}}
\newdimen\Tdim
\def\Tspan#1{{\setbox0=\hbox{#1}%
\Tdim\ht0\advance\Tdim\dp0\advance\Tdim.55ex\rule[-\dp0]{0pt}{\Tdim}\box0}}

\begin{document}

\begin{flushright}
YITP-22-10
\end{flushright}
\medskip
\renewcommand{\thefootnote}{\fnsymbol{footnote}}

\begin{center}
{\large\bf
Covariant BRST Quantization of Unimodular Gravity I \\
\vs{3}
--- Formulation with antisymmetric tensor ghosts ---
}
\vs{10}

{\large
Taichiro Kugo,$^{1,}$\footnote{e-mail address: kugo@yukawa.kyoto-u.ac.jp}
Ryuichi Nakayama,$^{2,}$\footnote{e-mail address: nakayama@particle.sci.hokudai.ac.jp}
and
Nobuyoshi Ohta\,$^{3,4,}$\footnote{e-mail address: ohtan@ncu.edu.tw}
} \\
\vs{5}

$^1$
{\em Yukawa Institute for Theoretical Physics, Kyoto University, Kyoto 606-8502, Japan}
\vs{3}

$^2$
{\em Division of Physics, Graduate School of Science,
Hokkaido University, Sapporo 060-0810, Japan}
\vs{3}

$^3$
{\em Department of Physics, National Central University, Zhongli, Taoyuan 320317, Taiwan}

and

$^4$
{\em Research Institute for Science and Technology,
Kindai University, Higashi-Osaka, Osaka 577-8502, Japan
}

\vs{5}
{\bf Abstract}
\end{center}

Unimodular gravity (UG) is an interesting theory that may explain why the cosmological constant is extremely
small, in contrast to general relativity (GR). The theory has only the transverse diffeomorphism invariance and
this causes a lot of debate as to the equivalence of UG to GR in the covariant quantization.
We study the covariant BRST quantization of UG by gauge fixing only the transverse diffeomorphism
and show that the remaining physical degrees of freedom are two, the same number as GR.
This is achieved by using antisymmetric tensor ghost fields which automatically satisfy the transverse condition
without nonlocal projection operator. The theory exhibits the ghosts for ghosts phenomenon, which
requires further gauge fixing and introduction of more ghosts.
We identify the BRST quartet structure
among the various fields and single out the remaining
physical degrees of freedom.

\renewcommand{\thefootnote}{\arabic{footnote}}
\setcounter{footnote}{0}

\newpage
\section{Introduction}

The recent cosmological observations have confirmed the current Universe is undergoing accelerated
expansion~\cite{AC1,AC2,AC3}.
The simplest way to realize this is to assume the existence of tiny vacuum energy or cosmological constant.

However, it is difficult to understand how and why such a small cosmological constant exists.
The problem gets its urgency if we consider the following classical and quantum aspects.
The classical aspect of the problem is that starting from the very early Universe  when the temperature is
extremely high, the Universe  would be in a symmetric phase, and when it makes transitions to the present state
including the electroweak symmetry breaking, huge vacuum energy would be generated.
This huge vacuum energy must be somehow canceled.
Moreover, even if this could be canceled, there remains a quantum problem.%
\footnote{It should, however, be kept in mind that
there is no clear distinction between classical and quantum vacuum energies.
The ``classical'' vacuum energy of the scalar potential has its origin in
the quantum dynamics causing the spontaneous symmetry
breaking~\cite{Kugo:2020ybc}.}
The vacuum fluctuations in quantum field
theory also induce a vacuum energy. The contribution of quantum fluctuations in known fields up to 300 GeV gives
a vacuum energy density of order (300 GeV)$^4$. This is vastly bigger than the observed dark energy density
$(3 \times10^{-3}\mbox{ eV})^4$ by a factor of order $10^{56}$.

Considering the huge size of the cosmological constant, it may be a first approximation to consider
a mechanism that the cosmological constant vanishes naturally.
In the standard gravity theory of general relativity (GR), there is no way to cancel the vast vacuum energy
except by fine tuning the cosmological constant already present in the theory.
But this is quite unnatural, if not impossible.
The unimodular gravity (UG) is one of such theories which may explain why the cosmological constant is
zero~\cite{BD1,BD2,Unruh,HT,EVMU,NV}.
This theory can be formulated by imposing the constraint that the determinant
of the metric is a fixed volume form in the general relativity:
\bea
S_{\rm UG}= Z_N \int d^4 x \left[ \sqrt{-g} (R-2\Lambda)
+ \la'(\sqrt{-g}-\omega)\right],
\label{eq:UGaction1}
\eea
where $Z_N= 1/(16\pi G_N)$ with $G_N$ being the Newton constant, and $\lambda'$ is a Lagrange multiplier
field to impose the constraint
\bea
\sqrt{-g}=\omega,
\label{unimodularity}
\eea
with $\omega$ being a fixed volume form.
We can shift $\la'$ to absorb the cosmological constant $\Lambda$ and obtain, up to a constant term,
\bea
S_{\rm UG}= Z_N \int d^4 x \left[ \sqrt{-g} R + \la(\sqrt{-g}-\omega)\right],
\label{eq:UGaction2}
\eea
where
\bea
\lambda=\lambda'-2\Lambda.
\eea

Making the variation of the action~\p{eq:UGaction1} or \p{eq:UGaction2} with respect to the metric, and
then eliminating the Lagrange multiplier, we can only derive the traceless part of the Einstein equation,
irrespective whether there is a cosmological constant or not.
Using the Bianchi identity together with the conservation of the energy-momentum tensor
in the presence of matter fields, we can recover the Einstein equation with a cosmological constant,
which now appears as an integration constant. The important point is that the cosmological constant has nothing to do
with the constant in the action. The cosmological constant is determined depending on the boundary condition,
and the above huge vacuum energies, classical or quantum, do not affect it.

Now since the UG could be regarded as just a partially gauge-fixed theory of GR, one would expect that
it is equivalent to GR, but there has been a lot of debate on the
equivalence~\cite{Smolin,FG,Eichhorn2013,Saltas,PS,Alvarez0,Alvarez1,BOT,Percacci2017,DOP,GM,HS,BP,DMPP,B,N}.
This problem has been discussed in various formulations of UG, most notably by Hamiltonian analysis.
For example, in~\cite{HT}, the authors find three primary and three secondary constraints which are of the first class.
These correspond to the transvserse diffeomorphism (TDiff) invariance, and eliminate 6 degrees of freedom (dofs).
Very interestingly, they find that there is a tertiary constraint of the first class, and this serves to eliminate
another dof. Thus we have originally $9$ dofs in the Hamiltonian formulation
because of the unimodular constraint, and $(6+1)=7$ dofs should be subtracted
owing to the first class constraints.
This leaves only $(9-7)=2$ dofs corresponding to the two physical graviton modes in the theory as in GR.

However it is more convenient to have a covariant quantization of the theory in order to carry out
covariant calculations.
In the standard BRST formulation of GR in the de Donder (or, harmonic) gauge, $\partial_\nu\tilde g^{\mu\nu}=0$ (
$\tilde g^{\mu\nu}:= \rtg g^{\mu\nu}$), we lift the full diffeomorphism in GR to
the BRST transformation by replacing the parameters $\varepsilon^\mu$
by the corresponding Faddeev-Popov (FP) ghosts $c^\mu$:
\bea
\d_{\B}  \tilde g^{\mu\nu} = -\rtg (\nabla^\mu c^\nu+ \nabla^\nu c^\mu)+ \tilde g^{\mu\nu}\nabla_\lambda c^\lambda,
\label{fdiff}
\eea
and impose the gauge-fixing condition by introducing pairs of antighosts $\bar c_\mu$ and Nakanishi-Lautrup (NL)
fields $b_\mu$,\cite{Nakanishi:1966zz}
which we call {\it multiplier BRST doublet transforming} as
\bea
\d_{\B}  \bar c_\mu= i b_\mu.
\label{adiff}
\eea
Generally for BRST doublets $(\bar c_\mu, b_\mu)$ transforming like this, we call
the $\bar c_\mu$ component the {\it BRST parent} and the $b_\mu$ component the {\it BRST daughter}.
The gauge-fixing term for the de Donder gauge condition $\pa_\mu\tilde g^{\mu\nu}=0$ and the corresponding
FP terms may be concisely written as~\cite{KU,Ohta2020}
\bea
{\cal L}_{\rm GF+FP} &=&
 -i \d_{\B}  \left[ \bar c_\mu \pa_\nu\tilde g^{\mu\nu}\right]
\nn &=&
b_\mu\pa_\nu \tilde g^{\mu\nu} -i \bar c_\mu \pa_\nu
\left[\rtg (\nabla^\mu c^\nu+ \nabla^\nu c^\mu)- \tilde g^{\mu\nu}\nabla_\lambda c^\lambda\right].
\eea
Four components in $g_{\mu\nu}$ become the BRST parents of the FP ghosts $c^\mu$, forming a vector BRST doublet.
The NL fields $b_\mu$, representing the other four components in $g_{\mu\nu}$ by equation of motion (EOM),
form another vector BRST doublet together with the FP antighosts $\bar c_\mu$.
Because these fields always come in this combination, forming a pair of BRST doublets,
this whole set of fields is called the {\it BRST quartet} and completely decouple
from the physical subspace~\cite{KO1977}.
These four sets of ghosts and antighosts leave $10-8=2$ physical dofs,
the two transverse massless spin-2 states.

In contrast, the problem is not so simple in UG, because the action~\p{eq:UGaction2} is not invariant under
the full diffeomorphism, but only under the TDiff (or, volume-preserving diffeomorphism):
\bea
\d g_{\mu\nu} = \nabla_\mu\e_\nu^{\T} +\nabla_\nu\e_\mu^{\T}, \qquad
\nabla^\mu\e_\mu^{\T}=0.
\eea
Thus we have only three sets of ghosts and antighosts, and this leaves $10-6=4$ dofs.
We also have the unimodularity condition~\p{unimodularity}.
However, this does not appear to introduce additional ghosts because the Lagrange multiplier field $\la$ is
a BRST singlet. It eliminates only one more dof, and we still have 3 dofs. How do we get rid of another dof?

In our previous paper~\cite{KNO}, with the perspective of getting insight into the covariant BRST quantization,
we have given a new formulation of BRST quantization of GR in the unimodular gauge with the gauge
condition~\p{unimodularity}. Unfortunately the formulation was not very successful for the quantization of UG.
The fundamental reason is that we did not properly gauge fix only the TDiff.
There is a paper discussing the covariant BRST quantization of UG~\cite{Alvarez1} by promoting
the theory with Weyl invariance. However it involves nontrivial nonlocal projection operators
and introduces a multitude of ghosts and antighosts whose origin is not easy to understand.

Here we would like to give covariant local BRST quantization of UG by just gauge fixing the TDiff
without introducing nonlocal projectors.
It has long been known in supergravity\cite{deWit:1980lyi,Sohnius:1982fw,Kugo:1982cu}
that a vector subject to transverse constraint can be expressed by an unconstrained antisymmetric tensor even in
the curved space-time.
We gauge fix TDiff by using such antisymmetric tensor fields,
which automatically satisfy the transverse condition. It turns out that after the gauge fixing,
the ghost kinetic terms have additional gauge invariance which has to be fixed, leading to additional ghosts
and antighosts. This is a phenomenon known as ghosts for ghosts since
the BRST quantization of antisymmetric tensor
gauge fields~\cite{Townsend:1979hd,Kimura:1980zd,HKO}.
We have to continue this process until there remains no more gauge invariance.
In this way, we find that we also have to introduce many ghosts, but our formulation is transparent because
the reason why these ghosts are necessary is clear.
We find that most of the dofs cancel out, leaving 2 dofs, the same number as GR.
Our important discovery is that the Lagrange multiplier $\lambda$ is identified with a BRST daughter by field equation,
forming a BRST quartet, and other modes in the ghost sector all cancel out.
The precise structure of the quartets is complicated due to the existence of multipole fields,
which will be clarified below.
The above question of how the unimodular constraint introduces additional ghosts is resolved in this way.

This paper is organized as follows.
In sect.~\ref{brstq}, we start with the gauge-fixing only TDiff
using antisymmetric rank-2 tensor ghosts and antighosts.
We find that the ghost kinetic term has additional gauge invariance,
which must be fixed, and this introduces secondary antisymmetric rank-3 tensor ghosts, whose kinetic term
has further gauge invariance. We go on to gauge fix it by introducing further tertiary ghosts of antisymmetric
rank-4 tensor. The process ends at this stage.
To simplify the following discussions, we transform these antisymmetric tensor fields to their duals.
In order to check that we have fixed all the gauge invariance, in sect.~\ref{brstq.2},
we calculate the resulting propagators on the flat backgrounds for simplicity.
The existence of the propagator proves that we gauge fix all the invariance.
To study the spectrum in the theory, we examine the EOMs at the linear order in sect.~\ref{eom}.
It turns out that some of them contain not only simple-pole fields, but also dipoles and tripoles.
In sect.~\ref{counting}, we identify which fields represent independent modes and how most of the fields fall
into the BRST quartets.
In sect.~\ref{metric}, we further clarify the metric structures of the BRST quartets and show that
there remain only 2 physical dofs in the theory.
In sect.~\ref{discussions}, we summarize our results and conclude the paper with some discussions.
We relegate some technical details to the Appendix, where we discuss the covariant divergence of antisymmetric tensors.

\section{BRST quantization of unimodular gravity }
\label{brstq}

In this section, we covariantly quantize the UG based on the BRST invariance.


In UG, we have only invariance under TDiff given by
\bea
\d g^{\mu\nu} = -\nabla^\mu\epsilon^\nu_{\T} -\nabla^\nu\epsilon^\mu_{\T}, \qquad
\nabla_\mu\epsilon^\mu_{\T} =0,
\label{tdiff}
\eea
which keeps the unimodularity condition $\sqrt{-g}=\omega$; indeed, the
transversality $\nabla_\mu\epsilon^\mu_{\T} =0$ is required because
\begin{equation}
\delta\rtg = -\frac12\rtg g_{\mu\nu}\delta g^{\mu\nu}
= -\frac12\rtg g_{\mu\nu}(-\nabla^\mu\epsilon^\nu_{\T} -\nabla^\nu\epsilon^\mu_{\T})
= \rtg\nabla_\mu\epsilon^\mu_{\T}.
\end{equation}
Then action (\ref{eq:UGaction2}) is invariant under the following BRST transformation:
\begin{eqnarray}
\delta_{\B}  g^{\mu\nu} & =& -\nabla^\mu c^\nu_{\T} -\nabla^\nu c^\mu_{\T}, 
\label{dbg} \\
\delta_{\B}  \lambda&=& 0, 
\label{dbl}
\end{eqnarray}
expressed in terms of diffeomorphism FP ghosts $c^{\mu}_{\T}$,
which satisfies a transversality condition:
\begin{equation}
\nabla_\mu c^\mu_{\T}=0.  
\label{trans}
\end{equation}
Solution to this transversality condition is in general nonlocal.
It is, however, actually known\cite{deWit:1980lyi,Sohnius:1982fw,Kugo:1982cu}
how to express the quantity subject to the
transverse constraint in terms of an unconstrained variable without
introducing any nonlocality.
Generally, for any totally antisymmetric contravariant tensor
$a^{\mu\nu\rho_1 \cdots\rho_n}$ of
rank-$(n+2)$ $(0\leq n\leq d-2)$, an identity holds:
\begin{equation}
\nabla_\mu\nabla_\nu a^{\mu\nu\rho_1 \cdots\rho_n}=0.
\label{eq:doubleDiv}
\end{equation}
Clearly the same form of identity also holds for the covariant
totally antisymmetric tensor $A_{\mu\nu\rho_1 \cdots\rho_n}$,
\begin{equation}
\nabla^\mu\nabla^\nu A_{\mu\nu\rho_1 \cdots\rho_n}=0,
\label{eq:doubleDiv2}
\end{equation}
since the covariant derivative commutes with the metric tensor.
These are explained in the Appendix.

So the transverse vector ghost $c^\mu_{\T}$ can generally be expressed in terms
of an antisymmetric rank-2 tensor ghost $c^{\mu\nu}$ as
\begin{equation}
c^\mu_{\T} = \nabla_\nu c^{\nu\mu},
\label{eq:1by2}
\end{equation}
whose covariant divergence vanishes by Eq.~(\ref{eq:doubleDiv}) for $n=0$.
Of course, in $d$-dimensional space-time,
$c^{\nu\mu}$ has $d(d-1)/2$ components larger than $d-1$ of $c^\mu_{\T}$
(for $d>2$), so that, given $c^\mu_{\T}$, the corresponding $c^{\mu\nu}$ is not
unique but has an ambiguity, which will become a gauge invariance yielding
the ghost for ghost~\cite{Townsend:1979hd,Kimura:1980zd,HKO}.
The ambiguity (or the invariance) in $c^{\nu\mu}$ in
Eq.~(\ref{eq:1by2}) is given by
the term of the form $\nabla_\rho c^{\rho\nu\mu}$ in terms of the rank-3 antisymmetric tensor,
whose covariant divergence vanishes again by Eq.~(\ref{eq:doubleDiv}) for $n=1$. The rank-3 tensor $c^{\rho\nu\mu}$
has an invariance of rank-4 antisymmetric tensor $c^{\sigma\rho\nu\mu}$, and so on,
ending at rank-$d$ antisymmetric tensor. Since the invariance at each step
gives a redundancy of the previous step tensor,
we can count the independent field components as
\begin{equation}
{}_dC_2-{}_dC_3+\cdots+(-1)^d{}_dC_d
= (1-1)^d -({}_dC_0-{}_dC_1) = -1+d.
\end{equation}
This exactly reproduces the independent component number of the
starting transverse vector $c^\mu_{\T}$!

This sequence of redundancy $=$ gauge invariance will appear in the following
gauge-fixing ghost Lagrangian, and also shows up in the
BRST transformation of those tensor ghosts which we now consider.

The nilpotency requirement of the BRST transformation on the metric field $g^{\mu\nu}$,
$\delta_{\B} (\delta_{\B}  g^{\mu\nu})=0$, is well known to determine the BRST transformation
of the ghost field as
\begin{equation}
\delta_{\B}  c^\mu_{\T} = c^\nu_{\T}\partial_\nu c^\mu_{\T}
= c^\nu_{\T}\nabla_\nu c^\mu_{\T} = \nabla_\nu\big(c_{\T}^\nu c^\mu_{\T}\big).
\label{eq:BRScmu}
\end{equation}
Here the second equality follows from
\begin{equation}
c^\nu_{\T}\nabla_\nu c^\mu_{\T} =
c^\nu_{\T}\bigl(\partial_\nu c^\mu_{\T} + \Gamma^\mu_{\nu\lambda}c^\lambda_{\T}\bigr),
\end{equation}
in which the covariantization term vanishes since $\Gamma^\mu_{\nu\lambda}$ is
$\nu$-$\lambda$ symmetric while $c^\nu_{\T}c^\lambda_{\T}$ is antisymmetric.
The third equality is due to the transversality $\nabla_\nu c_{\T}^\nu=0$.

Now consider how the BRST transformation of the rank-2 tensor ghost
$c^{\mu\nu}$ is determined from Eq.~(\ref{eq:1by2}).
First, noting the formula (\ref{eq:A1}) in the Appendix
\begin{equation}
\nabla_\mu a^{\mu\nu_1\cdots\nu_n} =
\rtg^{-1}\partial_\mu(\rtg a^{\mu\nu_1\cdots\nu_n}),
\nonumber
\end{equation}
valid for contravariant antisymmetric tensors and the BRST invariance of $\rtg$,
we see that the BRST transformation $\delta_{\B} $ is commutative with the
covariant divergence operator
$\nabla_\mu$ acting on $a^{\mu\nu_1\cdots\nu_n}$; i.e.,
\begin{equation}
\delta_{\B} (\nabla_\mu a^{\mu\nu_1\cdots\nu_n}) =
\nabla_\mu(\delta_{\B} a^{\mu\nu_1\cdots\nu_n}) .  \label{Bcov}
\end{equation}
Performing the BRST transformation of both sides of
Eq.~(\ref{eq:1by2}) and using Eq.~(\ref{eq:BRScmu}), we find
\begin{equation}
\nabla_\nu\bigl(\delta_{\B} c^{\nu\mu}- c^\nu_{\T}c^\mu_{\T}\bigr)=0 .
\end{equation}
This implies that the quantity inside the bracket is a transverse
antisymmetric rank-2 tensor, which must be a covariant divergence of
a rank-3 antisymmetric tensor $d^{\rho\nu\mu}$.
So we find that the BRST transformation law for $c^{\nu\mu}$ should be\footnote{%
All the ghost fields here and below are taken to be hermitian and the imaginary
factor $i$ here is put so as to make the BRST transformed field
$\delta\phi:= \lambda\,\delta_\B \phi$ hermitian when $\phi$ is hermitian.
Note that the transformation parameter $\lambda$ is an anti-hermitian
Grassmann-odd number, so that, in QED, for instance, the usual real gauge
transformation parameter $\theta$ is consistently replaced by
the transformation parameter $\lambda$ times the hermitian FP
ghost $c$; indeed, $(\lambda c)^\dagger = c^\dagger \lambda^\dagger =
c (-\lambda) = \lambda c$.
}
\begin{equation}
\delta_{\B} c^{\nu\mu} = c^\nu_{\T}c^\mu_{\T} + i\nabla_\rho d^{\rho\nu\mu} .
\label{eq:BRScnumu}
\end{equation}
This field $d^{\rho\nu\mu}$ is a hermitian boson carrying double ghost number $N_\FP=+2$
and denotes the ghost for ghost corresponding to the gauge
transformation of $c^{\mu\nu}$ under which the ``field strength''
$c^\mu_{\T} = \nabla_\nu c^{\nu\mu}$ is invariant.

As stated above, this sequence of gauge invariance continues.
Requiring the nilpotency of BRST transformation on the field $c^{\nu\mu}$
determines the BRST transformation of this ghost for ghost field $d^{\rho\nu\mu}$.
Performing BRST transformation of Eq.~(\ref{eq:BRScnumu}) and noting that
\begin{align}
\delta_{\B} (c^\nu_{\T}c^\mu_{\T})
&=\delta_{\B} c^\nu_{\T}\cdot c^\mu_{\T} -c^\nu_{\T} \cdot\delta_{\B}  c^\mu_{\T}
=c^\rho_{\T}\nabla_\rho c^\nu_{\T} \cdot c^\mu_{\T} -c^\nu_{\T} \cdot c^\rho_{\T}\nabla_\rho c^\mu_{\T}  \nn
&=c^\rho_{\T}\nabla_\rho\bigl(c^\nu_{\T} c^\mu_{\T}\bigr)
=\nabla_\rho\bigl(c^\rho_{\T}c^\nu_{\T} c^\mu_{\T}\bigr),
\end{align}
we find that the nilpotency of the BRST transformation leads to
\begin{equation}
\nabla_\rho\bigl(i\delta_{\B} d^{\rho\nu\mu} + c^\rho_{\T}c^\nu_{\T}c^\mu_{\T}\bigr)=0 ,
\end{equation}
which implies that the BRST transformation of $d^{\rho\nu\mu}$ is given in terms
of a rank-4 antisymmetric tensor ghost field $t^{\sigma\rho\nu\mu}$ in the form:
\begin{equation}
\delta_{\B} d^{\rho\nu\mu}= i c^\rho_{\T}c^\nu_{\T}c^\mu_{\T} - \nabla_\sigma t^{\sigma\rho\nu\mu}.
\label{eq:BRSdrnm}
\end{equation}
In the same way,
the nilpotency requirement on $d^{\rho\nu\mu}$ determines the BRST transformation
law of $t^{\sigma\rho\nu\mu}$ in the form in the $d=4$ case
\begin{equation}
\delta_{\B}  t^{\sigma\rho\nu\mu}= ic^\sigma_{\T} c^\rho_{\T}c^\nu_{\T}c^\mu_{\T} .
\end{equation}
There is no further invariance since there is no rank-5 totally antisymmetric tensors in $d=4$.

We emphasize that solely the requirement of the nilpotency of the BRST transformation completely determines
the BRST transformations of the ghost fields and clarifies that there are additional gauge invariances.

Now let us start the BRST quantization of this UG system.
Since we have only TDiff gauge symmetry, we can impose only
$d-1=3$ gauge-fixing conditions.
We prepare transverse {\it contravariant} vector multiplier BRST doublet:
\begin{equation}
\delta_{\B}  \bar c^\mu_{\T} = i b^\mu_{\T},  \qquad \delta_{\B}  b^\mu_{\T}=0,
\label{eq:1stMultiplierDoublet}
\end{equation}
subject to transversality conditions:
\begin{equation}
\nabla_\mu\bar c^\mu_{\T} = \nabla_\mu b^\mu_{\T}=0.
\end{equation}
Here $b^\mu_{\T}$ is the NL field which plays the role of
gauge-fixing multiplier and $\bar c^\mu_{\T}$ is its BRST parent, so may be called
a pre-multiplier. We generally call such a pair of fields,
the NL field (multiplier) and its BRST parent (pre-multiplier),
the {\it multiplier BRST doublet}.
We need to introduce such a multiplier BRST doublet for fixing each gauge invariance.

By using this multiplier BRST doublet (\ref{eq:1stMultiplierDoublet})
as the first step of gauge fixing,
we take the following gauge-fixing (GF) plus corresponding FP ghost  Lagrangian,
which corresponds to the de Donder gauge in GR,  $\partial_\nu\tilde g^{\nu\mu}=0$
for $\tilde{g}^{\mu\nu} :=\sqrt{-g} g^{\mu\nu}$~\cite{KU,Ohta2020}:
\begin{align}
{\cal L}_{\rm GF+FP,1} &= -i\d_{\B}  \bigl[g_{\mu\nu} \bar c^\nu_{\T} \pa_{\la} \tilde{g}^{\la\mu}
\bigr]
\nn
&= g_{\mu\nu}b^\nu_{\T} \partial_{\lambda}\tilde{g}^{\lambda\mu}-i(\nabla_{\mu}c^{\T}_{\nu}
+\nabla_{\nu}c^{\T}_{\mu})\bar c^\nu_{\T} \pa_{\la} \tilde{g}^{\la\mu}
 -ig_{\mu\nu}\bar{c}_{\T}^{\nu} \pa_{\la}
\bigl[\sqrt{-g} (\nabla^{\la}c_{\T}^{\mu}+\nabla^{\mu}c^{\la}_{\T})\bigr].
\label{gfaction0}
\end{align}
Because the multiplier field and the gauge-fixing function are both contravariant vectors, it is necessary to
introduce some covariant tensor to contract the indices. We use the full metric $g_{\mu\nu}$ to contract the indices.
Note also that the tensor indices
are raised or lowered by the full metric $g^{\mu\nu}, g_{\mu\nu}$ as usual;
e.g.,
$c_\mu^{\T} := g_{\mu\nu}c_{\T}^\nu$.

As the transverse ghost $c^\mu_{\T}$ has been expressed
in terms of an unconstrained antisymmetric tensor $c^{\nu\mu}$
as $\nabla_\nu c^{\nu\mu}$ in Eq.~(\ref{eq:1by2}), the multiplier
BRST doublet fields can also be written in the form
\begin{equation}
\bar c^\mu_{\T} = \nabla_\nu\bar c^{\nu\mu}, \qquad
b^\mu_{\T} = \nabla_\nu b^{\nu\mu},
\label{eq:barCBnumu}
\end{equation}
in terms of antisymmetric contravariant tensor fields $\bar c^{\nu\mu}$ and $b^{\nu\mu}$.
Just the same as for the above ghost case, here also exist
ambiguities in choosing $\bar c^{\nu\mu}$ for $\bar c^\mu_{\T}$ and $b^{\nu\mu}$ for
$b^\mu_{\T}$, which will also show up as the gauge invariance of the rewritten
Lagrangian. This ambiguity also exists when we rewrite the BRST transformation
law $\delta_{\B}  \bar c^\mu_{\T} = i b^\mu_{\T}$ into the relation between
$\delta_{\B} \bar c^{\nu\mu}$ and $b^{\nu\mu}$, but we fix it by defining the
$ib^{\nu\mu}$ as the BRST daughter of $\bar c^{\nu\mu}$:
\begin{equation}
\delta_{\B}  \bar c^{\nu\mu}= ib^{\nu\mu} .
\label{eq:dBcnumu}
\end{equation}
Note that Eq.~(\ref{eq:dBcnumu}) can be consistent with
$\delta_{\B}  \bar c^\mu_{\T} = i b^\mu_{\T}$ under the definition (\ref{eq:barCBnumu}),
thanks to the commutativity (\ref{Bcov}) of $\delta_\B$ and covariant divergence $\nabla_\mu$ which holds
only for the {\it contravariant tensor}.
This is the reason why we have introduced the multiplier BRST doublet
$\delta_{\B}  \bar c^\mu_{\T} = i b^\mu_{\T}$ as a contravariant vector despite the fact that
a covariant vector multiplier might seem more natural. Note also that
$\delta_{\B}  \bar{c}_{\mu}^{\T} \neq ib_{\mu}^{\T}$
(for
$\bar{c}_{\mu}^{\T}:= g_{\mu\nu}\bar{c}^{\nu}_{\T}$,
$b_{\mu}^{\T}:= g_{\mu\nu} b^{\nu}_{\T}$ )
and $\delta_{\B}  \bar{c}_{\nu\mu} \neq ib_{\nu\mu}$ since
$g_{\mu\nu}$ is not BRST invariant.

Using these rank-2 antisymmetric tensor fields, $c^{\nu\mu}$,
$\bar c^{\nu\mu}$ and $b^{\nu\mu}$, the GF+FP ghost Lagrangian (\ref{gfaction0}) is rewritten as
\begin{align}
{\cal L}_{\rm GF+FP,1}
&= -i\d_{\B}  \bigl[ g_{\mu\nu} \nabla_\rho\bar c^{\rho\nu}\cdot \pa_{\la} \tilde{g}^{\la\mu}
\bigr]
\nn
&=g_{\mu\nu} \nabla_\rho b^{\rho\nu}\cdot \partial_{\lambda}\tilde{g}^{\lambda\mu}
-i(\nabla_{\mu}\nabla^{\rho}c_{\rho\nu}+\nabla_{\nu}\nabla^{\rho}c_{\rho\mu})\nabla_{\sigma}
\bar c^{\sigma\nu} \pa_{\la} \tilde{g}^{\la\mu}  \nn
& \hspace{1em}{} -ig_{\mu\nu}\nabla_\rho\bar{c}^{\rho\nu} \cdot \pa_{\la}
\bigl[ \sqrt{-g} (\nabla^{\la}\nabla_\rho c^{\rho\mu}+\nabla^{\mu}\nabla_\rho c^{\rho\la}) \bigr].
\label{gfaction1}
\end{align}

As announced a few times in the above, this ghost Lagrangian has gauge invariance under transformations
with rank-3 totally antisymmetric parameters $\varepsilon^{\rho\nu\mu},
\ \bar\varepsilon^{\rho\nu\mu}$ and $\t^{\rho\nu\mu}$~\cite{HKO}:
\begin{align}
\delta c^{\nu\mu} &= \nabla_\rho\varepsilon^{\rho\nu\mu},
\label{eq:GT1-1}\\
\delta\bar c^{\nu\mu} &= \nabla_\rho\bar\varepsilon^{\rho\nu\mu},
\label{eq:GT1-2}\\
\delta b^{\nu\mu} &= \nabla_\rho\theta^{\rho\nu\mu},
\label{eq:GT1-3}
\end{align}
since this Lagrangian (\ref{gfaction1}) depends on these fields
only through
$c_{\T}^\mu=\nabla_\nu c^{\nu\mu}, \ \bar c^\mu_{\T}=\nabla_\nu\bar c^{\nu\mu}$ and
$b^\mu_{\T}=\nabla_\nu b^{\nu\mu}$, which
are invariant under these gauge transformations.
Note that the first gauge invariance (\ref{eq:GT1-1}) is already lifted in our
BRST transformation (\ref{eq:BRScnumu}) with the ghost for ghost field $d^{\rho\nu\mu}$,
and the second one (\ref{eq:GT1-2}) is included as a part of the multiplier
BRST transformation $\delta_{\B}  \bar c^{\nu\mu} =ib^{\nu\mu}$.
In order to fix the former two gauge invariances (\ref{eq:GT1-1}) and
(\ref{eq:GT1-2}), we take the following gauge-fixing conditions and introduce
the corresponding multiplier BRST doublets to impose them:
\begin{eqnarray}
\begin{array}{ccc}
\text{gauge-fixing cond.}  &   :      &  \text{multiplier BRST doublet} \\
\nabla^{[\rho}c^{\nu\mu]} = 0  &  :& (\bar{d}_{\rho\nu\mu},\bar c_{\rho\nu\mu}  ),  \quad
\delta_{\B}  \bar{d}^{\rho\nu\mu} = \bar c^{\rho\nu\mu},  \\
\nabla^{[\rho}\bar c^{\nu\mu]} = 0  &:& ( b_{\rho\nu\mu},  c_{\rho\nu\mu} ), \quad
\delta_{\B}  b^{\rho\nu\mu} = c^{\rho\nu\mu}.
 \end{array}
\label{eq:2ndGF}
\end{eqnarray}
Here the BRST transformation rules are also presented in addition to the doublet fields.
Note that although $\bar{d}_{\rho\nu\mu}$ and $b_{\rho\nu\mu}$ are covariant tensors due to their roles
as Lagrange multipliers for gauge-fixing of contravariant tensors, their BRST transformation rules are defined
in terms of  their contravariant partners. As we will see shortly, this definition will lead to antisymmetric
tensor gauge symmetry in the GF+FP ghost terms and allows introduction of further higher-rank antisymmetric
tensor gauge fields.
The gauge-fixing condition on the $b^{\nu\mu}$ field for the third gauge invariance
(\ref{eq:GT1-3}) is not necessary, since $b^{\nu\mu}$ is the BRST daughter of
$\bar c^{\nu\mu}$ and its gauge-fixing terms will be created from the gauge-fixing condition for
 $\bar{c}_{\nu\mu}$ via BRST transformation. The gauge condition will not, however, coincide simply
with $\nabla^{[\rho}\, b^{\nu\mu]} = 0$.


Now we can write down the second step GF+ FP ghost terms as
\begin{align}
{\cal L}_{\rm GF+FP,2} 
&= \frac{i}{2} \rtg  \delta_{\B}  \big(
\bar{d}_{\rho\nu\mu}\nabla^\rho c^{\nu\mu}
+b_{\rho\nu\mu}\nabla^\rho\bar{c}^{\nu\mu} \big) \nn
&= \frac{i}{2} \rtg  \delta_{\B}  \big(
\bar{d}^{\rho\nu\mu}\nabla_\rho c_{\nu\mu}
+b^{\rho\nu\mu}\nabla_\rho\bar{c}_{\nu\mu} \big).
\label{gfactor21}
\end{align}
Here, in the second expression, the tensor indices have been raised and
lowered for convenience in evaluating the
BRST transformation in this Lagrangian.
We note the identity
(\ref{eq:covRotation}) in the Appendix
\begin{equation}
\nabla_{[\mu}A_{\rho_1\cdots\rho_p]}= \partial_{[\mu}A_{\rho_1\cdots\rho_p]},
\end{equation}
holding for any antisymmetric {\it covariant} tensor $A_{\rho_1\cdots\rho_p}$ with any
rank $p$, which implies, in particular, that
the covariant derivative there also commutes with the BRST transformation:
\begin{equation}
\delta_\B\nabla_{[\mu}A_{\rho_1\cdots\rho_p]}= \partial_{[\mu}\delta_\B A_{\rho_1\cdots\rho_p]}=
\nabla_{[\mu}\delta_\B A_{\rho_1\cdots\rho_p]}.
\end{equation}
Then, noting that this commutativity holds for the parts $\nabla_\rho c_{\nu\mu}$ and
$\nabla_\rho\bar c_{\nu\mu}$ since they are multiplied by totally antisymmetric
tensors $\bar{d}^{\rho\nu\mu}$ and $b^{\rho\nu\mu}$, respectively,
we can rewrite ${\cal L}_{\rm GF+FP,2}$ into
\begin{align}
 {\cal L}_{\rm GF+FP,2}
&=\frac{i}{2} \rtg \big[ \bar{c}^{\rho\nu\mu}\nabla_\rho c_{\nu\mu} - \nabla_{\rho}\bar{d}^{\rho\nu\mu}
 \cdot \delta_{\B}  c_{\nu\mu}
+  c^{\rho\nu\mu} \, \nabla_\rho\bar{c}_{\nu\mu} -\nabla_{\rho} b^{\rho\nu\mu} \cdot \delta_{\B}
 \bar{c}_{\nu\mu} \big]    .
\end{align}
Here partial integrations have been performed in the second and fourth terms.
By taking account of $c_{\nu\mu}=g_{\nu\sigma}g_{\mu\kappa}c^{\sigma\kappa}$ and a similar equation for
$\bar{c}_{\nu\mu}$, we finally obtain
\begin{align}
{\cal L}_{\rm GF+FP,2}
& =\frac{i}{2} \rtg \Bigl[ \bar{c}^{\rho\nu\mu} \, \nabla_\rho c_{\nu\mu} - \nabla_{\rho}\bar{d}^{\rho\nu\mu}
 \cdot \delta_{\B}  (g_{\nu\sigma}g_{\mu\kappa})\,c^{\sigma\kappa}-\nabla^{\rho}\bar{d}_{\rho\sigma\kappa} \cdot
  (c^{\sigma}_{\T}c^{\kappa}_{\T}+i\nabla_{\mu}d^{\mu\sigma\kappa}) \nn
&\hspace{4em}{} +   c^{\rho\nu\mu} \, \nabla_\rho\bar{c}_{\nu\mu}-\nabla_{\rho} b^{\rho\nu\mu}
 \cdot \delta_{\B}  (g_{\nu\sigma}g_{\mu\kappa})\,\bar{c}^{\sigma\kappa}-i  \nabla^{\rho}  b_{\rho\nu\mu}
 \cdot  b^{\nu\mu} \Bigr] .
\label{gfaction2}
\end{align}
The gauge-fixing condition induced on $b^{\nu\mu}$ for the third gauge symmetry (\ref{eq:GT1-3}) may be read off from
the last two terms in (\ref{gfaction2}) which contain $b_{\rho\nu\mu}$:
\begin{align}
-i\delta_\B \bigl(
\nabla^{[\rho}\, \bar{c}^{\nu\mu]} \bigr)
=
\nabla^{[\rho}\, b^{\nu\mu]}+ 2i\nabla^{[\rho}\delta_\sigma^\nu g^{\mu]\lambda}(\delta_\B g_{\lambda\kappa})
\bar{c}^{\sigma\kappa} =0.
\end{align}
Because $[\delta_{\B} , \nabla^{\rho}] \neq 0$,
the BRST transformation of $\nabla^{[\rho}\, \bar{c}^{\nu\mu]}=0$ does not coincide with
$\nabla^{[\rho}\, b^{\nu\mu]}=0$.

This action (\ref{gfaction2}) still has gauge invariance under the
transformations:
\begin{align}
&\delta d^{\rho\nu\mu}=  \nabla_\sigma\varepsilon^{\sigma\rho\nu\mu} ,    \label{eq:GT2-1} \\
&\delta\bar{d}^{\rho\nu\mu}= \nabla_\sigma\bar\varepsilon^{\sigma\rho\nu\mu} , \label{eq:GT2-2} \\
&\delta b^{\rho\nu\mu}= \nabla_\sigma\epsilon^{\sigma\rho\nu\mu} , \label{eq:GT2-3} \\
&\delta\bar{c}^{\rho\nu\mu}= \nabla_\sigma\bar\theta^{\sigma\rho\nu\mu} , \label{eq:GT2-4} \\
&\delta c^{\rho\nu\mu}= \nabla_\sigma\theta^{\sigma\rho\nu\mu} , \label{eq:GT2-5}
\end{align}
because this action depends on these fields only through  their covariant divergences like
$\nabla_\rho d^{\rho\nu\mu}$, if partial integration is performed in case it is necessary.
Here again, the first gauge transformation  (\ref{eq:GT2-1}) is already lifted
in our BRST transformation (\ref{eq:BRSdrnm}) with the ghost for ghost for ghost field
$t^{\sigma\rho\nu\mu}$. The second and third transformations for the BRST parent fields
$\bar{d}^{\rho\nu\mu}$ and $b^{\rho\nu\mu}$ are contained as parts of
the multiplier BRST transformation in Eq.~(\ref{eq:2ndGF}). Again, we need
not put gauge-fixing conditions on the BRST daughter fields for the fourth and fifth gauge
invariances (\ref{eq:GT2-4}) and (\ref{eq:GT2-5}).
We fix the former three gauge invariances (\ref{eq:GT2-1}) to (\ref{eq:GT2-3})
by the following gauge-fixing conditions and introduce the corresponding multiplier BRST doublets to impose them:
\begin{eqnarray}
\begin{array}{ccc}
\text{gauge-fixing cond.}  &   :      &  \text{multiplier BRST doublet} \\
\nabla^{[\sigma}d^{\rho\nu\mu]} = 0  &:& (\bar{t}_{\sigma\rho\nu\mu},\bar{d}_{\sigma\rho\nu\mu}),  \quad
\delta_{\B}  \bar{t}^{\sigma\rho\nu\mu} = i\bar{d}^{\sigma\rho\nu\mu}, \\
\nabla^{[\sigma}\bar{d}^{\rho\nu\mu]} = 0 &:& (c_{\sigma\rho\nu\mu},d_{\sigma\rho\nu\mu} ), \quad
\delta_{\B}  c^{\sigma\rho\nu\mu} = i d^{\sigma\rho\nu\mu}, \\
\nabla^{[\sigma}b^{\rho\nu\mu]} = 0  &:& (\bar{c}_{\sigma\rho\nu\mu}, b_{\sigma\rho\nu\mu}  ), \quad
\delta_{\B}  \bar{c}^{\sigma\rho\nu\mu} = i b^{\sigma\rho\nu\mu}.
\end{array}
\label{eq:3rdGF}
\end{eqnarray}

We can now write down the third step GF+FP ghost Lagrangian as
\begin{align}
&\hspace{-1.5em}{\cal L}_{\rm GF+FP,3} \nn
&= -i \delta_{\B}  \Bigl[ \rtg
\frac1{3!}\bigl[
-\bar{t}_{\sigma\rho\nu\mu}\bigl(\nabla^{\sigma}d^{\rho\nu\mu} + \frac{\alpha}{4} d^{\sigma\rho\nu\mu}\bigr)
+c_{\sigma\rho\nu\mu}\nabla^{\sigma}\bar{d}^{\rho\nu\mu}
+\bar{c}_{\sigma\rho\nu\mu}\nabla^{\sigma}b^{\rho\nu\mu}
\bigr] \Bigr] \nn
&=
-i \delta_{\B}  \Bigl[ \rtg \frac{1}{3!}
\bigl[
-\bar{t}^{\sigma\rho\nu\mu}\bigl(\nabla_{\sigma}d_{\rho\nu\mu} + \frac{\alpha}{4} d_{\sigma\rho\nu\mu}\bigr)
+c^{\sigma\rho\nu\mu}\nabla_{\sigma}\bar{d}_{\rho\nu\mu}
+\bar{c}^{\sigma\rho\nu\mu}\nabla_{\sigma}b_{\rho\nu\mu}
\bigr]\Bigr],
\label{gfaction31}
\end{align}
where, in the second line, we have raised and lowered the tensor indices
for convenience to compute the BRST transformation in this Lagrangian just
in the same way as performed before for (\ref{gfactor21}):
\begin{align}
{\cal L}_{\rm GF+FP,3}&=  \frac{1}{6} \rtg \Bigl[
- \bar{d}^{\sigma\rho\nu\mu}\nabla_{\sigma}d_{\rho\nu\mu}
-\frac{\alpha}{4}\bar{d}^{\sigma\rho\nu\mu}d_{\sigma\rho\nu\mu}
+i\nabla^{\sigma}\bar{t}_{\sigma\rho\nu\mu} \cdot \bigl(-\nabla_\lambda t^{\lambda\rho\nu\mu}
+ic^\rho_{\T}c^\nu_{\T}c^\mu_{\T}\bigr) \nn
&\hspace{4em}{}
+d^{\sigma\rho\nu\mu}\nabla_{\sigma}\bar{d}_{\rho\nu\mu}+ic^{\sigma\rho\nu\mu}\nabla_{\sigma}\bar{c}_{\rho\nu\mu}
+b^{\sigma\rho\nu\mu}\nabla_{\sigma}b_{\rho\nu\mu}
+i\bar{c}^{\sigma\rho\nu\mu}\nabla_{\sigma}c_{\rho\nu\mu} \nn
& \hspace{4em}{} +i \nabla_{\sigma} \bar{t}^{\sigma\rho\nu\mu} \cdot
\delta_\B (g_{\rho\kappa}g_{\nu\tau}g_{\mu\la})\,d^{\kappa\tau\la}
 -i\frac{\alpha}{4}\bar{t}^{\sigma\rho\nu\mu}
\delta_\B(g_{\sigma\kappa}g_{\rho\tau}g_{\nu\la}g_{\mu\chi})\,d^{\kappa\tau\la\chi} \nn
&\hspace{4em}{} -i\nabla_{\sigma}c^{\sigma\rho\nu\mu} \cdot
\delta_{\B} (g_{\rho\kappa}g_{\nu\tau}g_{\mu\la})\,\bar{d}^{\kappa\tau\la}
 -i\nabla_{\sigma}\bar{c}^{\sigma\rho\nu\mu}
\cdot \delta_{\B} (g_{\rho\tau}g_{\nu\kappa}g_{\mu\la})\,b^{\tau\kappa\la}
\Bigr]   .
\label{gfaction3}
\end{align}
Here we have introduced a gauge parameter $\alpha$ for later convenience.
We could have introduced more gauge parameters multiplied by such
(BRST daughter)$^2$ terms in our gauge-fixing actions (\ref{gfaction1}),
(\ref{gfaction2}) and (\ref{gfaction3}); we omitted them here other than
$\alpha$ since they are not useful for simplifying the structure of the propagators.
The last term on the third line vanishes identically, because it is proportional to $\delta_{\B}  \rtg$.
The gauge conditions for the fourth (\ref{eq:GT2-4}) and fifth (\ref{eq:GT2-5}) gauge symmetries
can be derived from the terms in (\ref{gfaction3}) which contain $c^{\sigma\rho\nu\mu}$ and
$\bar{c}^{\sigma\rho\nu\mu}$, respectively.


Now there remains no further invariance and we expect that the system is fully gauge fixed.
To avoid too many tensor suffices, we rewrite the tensor fields by their dual fields.
Generally in the curved space-time, it is convenient to define
the covariant antisymmetric tensor $A_{\mu_1\cdots\mu_q}$ (Hodge) dual to
a contravariant antisymmetric tensor $a^{\nu_1\cdots\nu_p}$ with $p+q=d$
by the relation Eq.~(\ref{eq:A7}):
\begin{equation}
\rtg a^{\mu_1\cdots\mu_p} =\pm(q!)^{-1}\varepsilon^{\mu_1\cdots\mu_p\nu_1\cdots\nu_q} A_{\nu_1\cdots\nu_q},
\end{equation}
or, by its inverse relation,
\begin{equation}
A_{\nu_1\cdots\nu_q} =  \mp (p!)^{-1}
\rtg a^{\mu_1\cdots\mu_p}
\varepsilon_{\mu_1\cdots\mu_p\nu_1\cdots\nu_q},
\end{equation}
(double sign in the same order). The point is that the $\rtg$ factor is attached to the contravariant tensor side.
So our sequence of ghost fields, $c^{\mu\nu}, \ d^{\mu\nu\rho}$ and $t^{\mu\nu\rho\sigma}$
are expressed by their dual fields $C_{\mu\nu},\ D_\mu,\ T$ (generally
denoted by the corresponding uppercase letters) as
\begin{align}
\rtg c^{\mu\nu} &=  (1/2)\varepsilon^{\mu\nu\rho\sigma} C_{\rho\sigma}  ,  \nn
\rtg d^{\mu\nu\rho}&=  \varepsilon^{\mu\nu\rho\sigma} D_\sigma,   \nn
\rtg t^{\mu\nu\rho\sigma}&= \varepsilon^{\mu\nu\rho\sigma} T .
\end{align}
The $1+2+3=6$ multiplier BRST doublets are expressed by their duals as
\begin{align}
& \rtg
\begin{pmatrix}
 \bar{c}^{\mu\nu} \\
b^{\mu\nu}
\end{pmatrix}
= -\frac12 \varepsilon^{\mu\nu\rho\sigma}
\begin{pmatrix}
\bar{C}_{\rho\sigma} \\
B_{\rho\sigma}
\end{pmatrix} ,
\qquad &
\rtg \begin{pmatrix}
\bar{t}^{\mu\nu\rho\sigma} \\
\bar{d}^{\mu\nu\rho\sigma}
\end{pmatrix}
= - \varepsilon^{\mu\nu\rho\sigma}
\begin{pmatrix}
\bar{T} \\
\bar{D}
\end{pmatrix} ,
\nn
&
\rtg
\begin{pmatrix}
\bar{d}^{\mu\nu\rho} \\
\bar{c}^{\mu\nu\rho}
\end{pmatrix}
= - \varepsilon^{\mu\nu\rho\sigma}
\begin{pmatrix}
\bar{D}_\sigma\\
\bar{C}_\sigma
\end{pmatrix} ,
\qquad &
\rtg
\begin{pmatrix}
c^{\mu\nu\rho\sigma} \\
d^{\mu\nu\rho\sigma}
\end{pmatrix}
= -\varepsilon^{\mu\nu\rho\sigma}
\begin{pmatrix}
C \\
D
\end{pmatrix} ,
\nn
&
\rtg
\begin{pmatrix}
b^{\mu\nu\rho} \\
c^{\mu\nu\rho}
\end{pmatrix}
= -\varepsilon^{\mu\nu\rho\sigma}
\begin{pmatrix}
B_\sigma\\
C_\sigma
\end{pmatrix} ,
\qquad &
\rtg
\begin{pmatrix}
\bar{c}^{\mu\nu\rho\sigma} \\
b^{\mu\nu\rho\sigma}
\end{pmatrix}
= -\varepsilon^{\mu\nu\rho\sigma}
\begin{pmatrix}
\bar{C} \\
B
\end{pmatrix} .
\end{align}
Furthermore $c^\mu_{\T}$, $\bar{c}^{\mu}_{\T}$ and $b^{\mu}_{\T}$ should be understood
to represent
\begin{align}
c^\mu_{\T}& = \nabla_\nu c^{\nu\mu} = -(2\rtg)^{-1} \varepsilon^{\mu\nu\rho\sigma} \partial_\nu C_{\rho\sigma} , \nn
\bar{c}^\mu_{\T}& = \nabla_\nu\bar{c}^{\nu\mu} =+ (2\rtg)^{-1} \varepsilon^{\mu\nu\rho\sigma}
 \partial_\nu\bar{C}_{\rho\sigma} , \nn
b^{\mu}_{\T} &= \nabla_\nu b^{\nu\mu} = + (2\rtg)^{-1} \varepsilon^{\mu\nu\rho\sigma} \partial_\nu B_{\rho\sigma} .
\label{eq:bTBb}
\end{align}

In terms of these dual field variables,
the GF+FP ghost Lagrangians (\ref{gfaction1}), (\ref{gfaction2}) and
(\ref{gfaction3}) are rewritten as
\begin{align}
\hs{-20}
{\cal L}_{\rm GF+FP,1}
&=-\frac{1}{2 \rtg} g_{\mu\nu} \varepsilon^{\rho\nu\sigma\tau} \nabla_\rho B_{\sigma\tau}\cdot
 \partial_{\lambda}\tilde{g}^{\lambda\mu} \nn
&   \hspace{1em}{} -\frac{i}{2(\rtg)^2} \varepsilon^{\chi\tau\rho\sigma}\varepsilon^{\nu\kappa\alpha\beta}
 g_{\chi(\mu}\nabla_{\nu)} \nabla_{\tau}C_{\rho\sigma} \cdot \nabla_{\kappa}\bar{C}_{\alpha\beta} \cdot
 \partial_{\la}\tilde{g}^{\la\mu} \nn
& \hspace{1em}{}+\frac{i}{2\rtg} g_{\mu\nu}\varepsilon^{\nu\rho\sigma\tau} \nabla_{\rho}\bar{C}_{\sigma\tau}
 \cdot \epsilon^{\kappa\alpha\beta(\mu} \partial_{\la}\nabla^{\la)}\nabla_{\kappa}C_{\alpha\beta}
,
\label{gfaction1P}
\end{align}
\begin{align}
{\cal L}_{\rm GF+FP,2}
&=
 \rtg \Big[ i\bar{C}^{\sigma}\nabla^\rho C_{\rho\sigma}
 +\frac{3i}{4\rtg}\varepsilon^{\kappa\mu\nu\la} \nabla^{\rho} \bar{D}^{\sigma}\cdot
 \nabla_{[\rho}C_{\kappa\sigma]}\cdot \nabla_{\mu}C_{\nu\la}\nn
&\hspace{4em}{}+
(\nabla^{\la}\bar{D}^{\sigma}-\nabla^{\sigma}\bar{D}^{\la})\cdot \nabla_{\la}D_{\sigma}
 -iC^{\sigma}\nabla^{\rho}\bar{C}_{\rho\sigma}
+B^{\sigma}\nabla^{\rho}B_{\rho\sigma}
 \nn
&\hspace{4em}{}-ig^{\mu\la}\big(\nabla^{\rho}\bar{D}^{\sigma}- \nabla^{\sigma}\bar{D}^{\rho}\big)
C_{\mu\sigma}\delta_{\B}  g_{\la\rho}     +ig^{\mu\la}\big(\nabla^{\rho}B^{\sigma}
- \nabla^{\sigma}B^{\rho}\big)\bar{C}_{\mu\sigma} \delta_{\B}  g_{\la\rho} \Big]
 ,  \label{gfaction2P}
\end{align}
\begin{align}
{\cal L}_{\rm GF+FP,3}
&=  \sqrt{-g} \Big[\bar{D}\nabla^{\mu}D_{\mu}+\alpha\bar{D}D +D \nabla_{\mu} \bar{D}^{\mu}
+iC\nabla_{\mu}\bar{C}^{\mu}+i \bar{C} \nabla_{\mu}C^{\mu} +B\nabla_{\mu}B^{\mu}  \nn
 &\hspace{3em}{}
-\frac{1}{4}\nabla^{\sigma}\bar{T} \cdot \nabla_{[\sigma}C_{\nu\mu]}
\big( \nabla^{\mu}C^{\la\rho}\cdot \nabla^{\nu}C_{\la\rho}-4\nabla_{\la}{C^{\mu}}_{\rho} \cdot \nabla^{\nu}C^{\la\rho}
\nn
&\hspace{3em}{} +2\nabla_{\la}{C^{\mu}}_{\rho}\cdot \nabla^{\la}C^{\nu\rho}
-2\nabla_{\la}{C^{\mu}{}_{\rho} \cdot\nabla^\rho} C^{\nu\la}\big) -i\nabla^{\mu}\bar{T} \cdot \nabla_{\mu}T
 \nn
& \hspace{3em}{}
+\bigl( i\nabla^{\mu}\bar{T} \cdot D^{\nu} +i\nabla^{\mu}C\cdot \bar{D}^{\nu}
+i\nabla^{\mu}\bar{C}\cdot B^{\nu} \bigr) \delta_{\B}  g_{\mu\nu}
\Big] .
\label{gfaction3P}
\end{align}
Here
\begin{align}
\delta_{\B}  g_{\mu\nu} &= -(\rtg)^{-1}g_{\la (\nu}\varepsilon^{\la\rho\sigma\tau}\nabla_{\mu)}\nabla_{\rho}C_{\sigma\tau},
\end{align}
is to be substituted in the above equations, and the brackets $( \ )$ and $[\ ]$
attached to the indices mean the weight 1 symmetrization and antisymmetrization,
respectively;
e.g., $A_{(\mu}B_{\nu)}=(1/2)(A_\mu B_\nu+A_\nu B_\mu)$.

The BRST transformations for the dual ghost fields are rewritten as follows:
\begin{align}
\delta_{\B}  C_{\mu\nu} &= -(1/2)\rtg \varepsilon_{\mu\nu\rho\sigma}c^\rho_{\T}c^\sigma_{\T}
 + i(\partial_\mu D_\nu-\partial_\nu D_\mu), \nn
\delta_{\B}  D_\mu&= (i/3!)\rtg \varepsilon_{\mu\nu\rho\sigma}c^\nu_{\T}c^\rho_{\T}c^\sigma_{\T} + \partial_\mu T, \nn
\delta_{\B}  T &=- (i/4!)\rtg \varepsilon_{\mu\nu\rho\sigma}c^\mu_{\T}c^\nu_{\T}c^\rho_{\T}c^\sigma_{\T}.
\end{align}
The BRST transformations for the dual fields of multiplier BRST doublets are trivial for covariant
tensors (with lower indices):
\begin{align}
&\delta_{\B}  \bar{C}_{\mu\nu}=i B_{\mu\nu}, \qquad
\delta_{\B}  \bar{D}_\mu=\bar{C}_\mu,
\qquad
\delta_{\B}  B_\mu= C_\mu, \nn
&\delta_{\B}  \bar{T}=i\bar{D}, \qquad \delta_{\B}  C=iD, \qquad \delta_{\B}  \bar{C}=i B.
\end{align}

\section{Propagators and equations of motion at linear order}

\subsection{Propagators}
\label{brstq.2}

Now the total Lagrangian of our UG system is given by
\begin{equation}
{\cal L}_{\rm UG} = \rtg R + \lambda(\rtg - \omega) +
{\cal L}_{\rm GF+FP,1} (\ref{gfaction1P}) +
{\cal L}_{\rm GF+FP,2} (\ref{gfaction2P}) +
{\cal L}_{\rm GF+FP,3} (\ref{gfaction3P})\,.
\label{eq:totalL}
\end{equation}
Let us check in detail if we get nonsingular fully gauge-fixed action on the flat background with $\omega=1$.
We introduce a fluctuation $h^{\mu\nu}$
of $\tilde{g}^{\mu\nu}:= \rtg g^{\mu\nu}$ around the flat metric
$\eta^{\mu\nu}$ defined by
\begin{equation}
\tilde{g}^{\mu\nu}=\eta^{\mu\nu}+h^{\mu\nu},
\end{equation}
and then to the linear order we have
\begin{equation}
g_{\mu\nu}=\eta_{\mu\nu}-h_{\mu\nu}+\frac{1}{2}\eta_{\mu\nu}h+\cdots, \qquad
\sqrt{-g}=
1+\frac12 h+\cdots.
\end{equation}
In what follows indices of the fields will be raised and lowered by using
$\eta^{\mu\nu}$ and $\eta_{\mu\nu}$,
respectively.
The quadratic terms in our total Lagrangian (\ref{eq:totalL})
are given by
\begin{eqnarray}
{\cal L}_{\rm UG}\Bigr|_{\rm quadr} &=&
{\cal L}_{N_{\FP}=0} + {\cal L}_{|N_{\FP}|=1} +    {\cal L}_{|N_{\FP}|=2} +{\cal L}_{|N_{\FP}|=3},  \nn
{\cal L}_{N_{\FP}=0}
&=&   \frac14 h_{\mu\nu}\square h^{\mu\nu}+\frac12 (\pa_\nu h^{\mu\nu})^2
-\frac18 h\square h +\frac12 \lambda h \nn
&& +\frac12 \e_{\mu\nu\rho\s} \pa^\nu B^{\rho\s} \pa_\la h^{\mu\la}
+ \pa_\nu B^{\nu\mu} B_\mu+B\pa^\mu B_\mu
, \nn
{\cal L}_{|N_{\FP}|=1}&=&
\frac{i}{2} \bar C^{\mu\nu} \square(\square C_{\mu\nu}+2 \pa_{[\mu}\pa^\rho C_{\nu]\rho})
+ i \bar C^\mu\pa^\nu C_{\nu\mu}+i\pa_\nu\bar C^{\nu\mu} C_\mu\nn
&&  -i \pa_\mu\bar C^\mu\cdot C +i \bar C \pa^\mu C_\mu
,\nn
{\cal L}_{|N_{\FP}|=2}&=&
 -\bar D^\mu(\square D_\mu-\pa^\nu\pa_\mu D_\nu)
+ \bar D\, \pa^\mu D_\mu+\pa_\mu\bar D^\mu\cdot D +  \a \, \bar D D
,\nn
{\cal L}_{|N_{\FP}|=3}&=&
 i \bar T\, \square T.
\end{eqnarray}
The coefficients of these give 2-point vertex matrix $\Gamma^{(2)\,ij}$,
the inverse of which multiplied by $i$ gives the
propagators; $i{\Gamma^{(2)}}^{-1}_{ij}=\VEV{ \T \phi_i\,\phi_j}$.

We start with $N_{\FP}=0$ sector. The 2-point vertex $\Gamma^{(2)}_{N_{\FP}=0}$ in momentum space is
\begin{eqnarray}
&&\hspace{-1em}\Gamma^{(2)}_{N_{\FP}=0} = \nonumber \\
&& \bordermatrix{
           & h_{\rho\sigma}  & B_{\rho\s} & B_\rho& B & \la  \cr
h_{\mu\nu} &
\begin{matrix}
\hs{-10}-p^2\,\Bigl[\frac12 P^{(2)\mu\nu,\rho\sigma}-\frac1{12}d^{\mu\nu}d^{\rho\sigma}
\\
-\frac14\left(d^{\mu\nu}e^{\rho\sigma}+e^{\mu\nu}d^{\rho\s}\right)
-\frac34e^{\mu\nu}e^{\rho\sigma}\Bigr]
\end{matrix}
   & \frac12 \varepsilon^{\a\b\rho\s}p_\b \d_\a^{(\mu} p^{\nu)} & 0 & 0 & \frac12\eta^{\mu\nu}
         \cr
B_{\mu\nu} & -\frac12 \varepsilon^{\mu\nu\b(\rho} p^{\s)} p_\b  & 0 
 &  - ip^{[\mu}\eta^{\nu]\rho}  &  0 & 0 \cr
B_\mu& 0 & -i \eta^{\mu[\rho}p^{\s]}  & 0 & -ip^\mu& 0 \cr
B & 0 & 0& i p^\rho& 0& 0 \cr
\la & \frac12 \eta^{\rho\sigma}  & 0 & 0 & 0 & 0 \cr
},\nn
\end{eqnarray}
with the projection operators
\def\II{{I\kern-2.5ptI}}
\begin{eqnarray}
&&d_{\mu\nu} = \eta_{\mu\nu}- \frac{p_\mu p_\nu}{p^2}, \qquad
e_{\mu\nu}= \frac{p_\mu p_\nu}{p^2},  \\
&& P^{(2)}_{\mu\nu,\rho\sigma}= \frac12\left(
d_{\mu\rho}d_{\nu\sigma}+d_{\mu\sigma}d_{\nu\rho}-\frac23 d_{\mu\nu}d_{\rho\sigma}\right),
\end{eqnarray}
which satisfy
\begin{eqnarray}
&&p^\mu d_{\mu\nu}=0,\quad d_{\mu\nu}\eta^{\mu\nu}=3,\quad e_{\mu\nu}\eta^{\mu\nu}=1, \\
&&
d_{\mu\alpha}d^{\alpha\nu}=d_\mu{}^\nu,\quad e_{\mu\alpha}e^{\alpha\nu}=e_\mu{}^\nu,
\quad d_{\mu\alpha}e^{\alpha\nu}=0, \\
&& P^{(2)}_{\mu\nu,\alpha\beta}d^{\alpha\beta}= 0,\quad
P^{(2)}_{\mu\nu,\alpha\beta}e^{\alpha\beta}= 0,\quad
P^{(2)}_{\mu\nu,\alpha\beta}P^{(2) \alpha\beta,\rho\sigma}= P^{(2)}_{\mu\nu}{}^{\rho\sigma}.
\end{eqnarray}

We can straightforwardly compute the inverse of the matrix,
${\Gamma^{(2)}}^{-1}_{N_{\FP}=0}$:
\begin{eqnarray}
&&\hspace{-1em}{\Gamma^{(2)}}^{-1}_{N_{\FP}=0}=
\frac1{-p^2}\times
\nonumber \\
&& \hs{-8}\bordermatrix{
           & h_{\rho\sigma}  & B_{\rho\s} & B_\rho& B & \la  \cr
h_{\mu\nu} &
\begin{matrix}
\Bigl[ 2 P^{(2)}_{\mu\nu,\rho\sigma}
-\frac13 d_{\mu\nu}d_{\rho\sigma} \hspace{4em}
\\
+(d_{\mu\nu}e_{\rho\sigma}+ e_{\mu\nu}d_{\rho\sigma})
-3 e_{\mu\nu}e_{\rho\sigma}\Bigr]
\end{matrix}
    & \displaystyle\frac{2}{p^2}\varepsilon_{\rho\s\la(\mu}\, p_{\nu)}p^\la & 0 & 0
&
\begin{matrix}
-p^2\times\hspace{1.8em}\\[-.4ex]
(d_{\mu\nu}-e_{\mu\nu})
\end{matrix}
\cr
B_{\mu\nu} & \displaystyle\frac{2}{p^2}\varepsilon_{\mu\nu\la(\rho}\, p_{\s)}p^\la  & 0  &
\hspace{-1em}-2i \eta_{\rho[\mu} p_{\nu]} &0 & 0 \cr
B_\mu& 0 & 2i\eta_{\mu[\rho} p_{\s]}  & 0 & ip_\mu& 0 \cr
B & 0 & 0 & -ip_\rho& 0 & 0 \cr
\la & -p^2 ( d_{\rho\sigma}-e_{\rho\sigma})  & 0 & 0 & 0 & 0 \cr
}. \nn
\label{eq:GammaInNF0}
\end{eqnarray}

The 2-point vertex $\Gamma^{(2)}_{|N_{\FP}|=1}$ in momentum space is
\begin{eqnarray}
\Gamma^{(2)}_{|N_{\FP}|=1} =
i \times\bordermatrix{
         &   C_{\rho\s} & C_\rho& C  \cr
\bar C^{\mu\nu} & \frac12 p^4\d^\rho_{[\mu} \d^\s_{\nu]}
-p^2 p^{\phantom{\nu}}_{[\mu}p_{\phantom{b}}^{[\rho}\d^{\s]}_{\nu]}
& -ip^{\phantom{\nu}}_{[\mu}\d^\rho_{\nu]} & 0 \cr
\bar C^\mu&  ip_{\phantom{b}}^{[\rho} \d^{\s]}_\mu&  0 & ip_\mu\cr
\bar C & 0 & ip^\rho& 0 \cr
},
\end{eqnarray}
the inverse of which is given by
\begin{equation}
{\Gamma^{(2)}}^{-1}_{|N_{\FP}|=1}=
\frac{i}{p^2}\times
\bordermatrix{
         &  \bar C^{\rho\s} & \bar C^\rho& \bar C  \cr
C_{\mu\nu} & \displaystyle\frac{2}{-p^2} \left[\d^{[\rho}_\mu\d^{\s]}_\nu
- \frac{2}{p^2} p^{\phantom{\nu}}_{[\mu}p_{\phantom{b}}^{[\rho} \d^{\s]}_{\nu]}
\right]
&  2ip^{\phantom{\nu}}_{[\mu}\d_{\nu]}^\rho& 0 \cr
C_\mu&  -2i p_{\phantom{b}}^{[\rho} \d^{\s]}_\mu&  0 & ip_\mu\cr
C & 0 & ip^\rho&0 \cr
}.
\label{eq:GammaInNF1}
\end{equation}

The 2-point vertex $\Gamma^{(2)}_{|N_{\FP}|=2}$ in momentum space is
\begin{eqnarray}
\Gamma^{(2)}_{|N_{\FP}|=2} =
\bordermatrix{
         &   D_\rho& D \cr
\bar D^\mu& p^2\d^\rho_\mu- p_\mu p^\rho& -ip_\mu\cr
\bar D  &  ip^\rho&  \a  \cr
},
\end{eqnarray}
the inverse of which is given by
\begin{equation}
{\Gamma^{(2)}}^{-1}_{|N_{\FP}|=2}=
\frac1{-p^2}\times
\bordermatrix{
         & \bar D^\rho& \bar D \cr
D_\mu& \displaystyle\left[-\d^\rho_\mu+(\a+1) \frac{p_\mu p^\rho}{p^2}
\right] & ip_\mu\cr
D &  -ip^\rho& 0  \cr
}.
\label{eq:GammaInNF2}
\end{equation}

Finally the 2-point vertex $\Gamma^{(2)}_{|N_{\FP}|=3}$ for $T$ and $\bar{T}$ in momentum space is given by
\begin{eqnarray}
\Gamma^{(2)}_{|N_{\FP}|=3} = -ip^2,
\end{eqnarray}
and the inverse of which is given by
\begin{equation}
{\Gamma^{(2)}}^{-1}_{|N_{\FP}|=3}=\frac{i}{p^2}.
\label{eq:GammaInNF3}
\end{equation}

Thus we have confirmed that the propagators exist and the system is fully gauge fixed.

\subsection{Equations of motion at linear order}
\label{eom}

Let us denote the quadratic part of the total action as $S$.
Classical EOMs to linear order are given as follows:
for $\alpha=-1$,
\begin{align}
\hbox{$N_\FP=0$ sector}\ & \nn
\frac{\delta S}{\delta\lambda} :\quad & h(\equiv\eta_{\mu\nu}h^{\mu\nu})=0,
\label{eq:N01} \\
\frac{\delta S}{\delta h^{\mu\nu}} :\quad & \frac{1}{2}\square h_{\mu\nu}-\partial_{(\mu}(h_{\nu)}
+b^{\T}_{\nu)})+\frac{1}{2}\lambda\eta_{\mu\nu}=0,
\label{eq:N02}\\
\frac{\delta S}{\delta B^{\rho\sigma}} :\quad &\frac{1}{2}\varepsilon_{\lambda\mu\rho\sigma}
\partial^{\lambda}\partial_{\nu}h^{\mu\nu}-\partial_{[\rho}B_{\sigma]}=0,
\label{eq:N03}\\
\frac{\delta S}{\delta B_{\mu}}:\quad & \partial_{\nu}B^{\nu\mu}-\partial^{\mu}B=0,
\label{eq:N04}\\[.5ex]
\frac{\delta S}{\delta B}:\quad & \partial^{\mu}B_{\mu}=0,
\label{eq:N05}\\[.5ex]
\hbox{$N_\FP=\pm1$ sector}\ &  \nn
\frac{\delta S}{\delta\bar{C}^{\mu}},\ \frac{\delta S}{\delta C_{\mu}} :\quad
& \partial^{\nu}C_{\nu\mu}+\partial_{\mu}C=0, \quad \partial_{\nu}\bar{C}^{\nu\mu}-\partial^{\mu}\bar{C}=0,
\label{eq:N11} \\
\frac{\delta S}{\delta\bar{C}^{\mu\nu}},\ \frac{\delta S}{\delta C_{\mu\nu}} :\quad
& \frac12 \square^2C_{\mu\nu}+\partial_{[\mu} C_{\nu]}=0,
\quad \frac12 \square^2\bar{C}^{\mu\nu}+\partial^{[\mu}\bar{C}^{\nu]}=0,
\label{eq:N12} \\
\frac{\delta S}{\delta\bar{C}},\ \frac{\delta S}{\delta C} :\quad & \partial^{\mu} C_{\mu}=0,
\quad \partial_{\mu}\bar{C}^{\mu}=0,
\label{eq:N13} \\[.5ex]
\hbox{$N_\FP=\pm2$ sector}\ &  \nn
\frac{\delta S}{\delta\bar{D}}, \ \frac{\delta S}{\delta D} :\quad
& \partial^{\mu}D_{\mu}=D,\quad \partial_{\mu}\bar{D}^{\mu}=\bar{D},
\label{eq:N21} \\
\frac{\delta S}{\delta\bar{D}^{\mu}},\ \frac{\delta S}{\delta D_{\mu}} :\quad
& \square D_{\mu}=0,\quad \square\bar{D}^{\mu}=0,
\label{eq:N22} \\[.5ex]
\hbox{$N_\FP=\pm3$ sector}\ &  \nn
\frac{\delta S}{\delta\bar{T}},\ \frac{\delta S}{\delta T} :\quad & \square T=0,\quad  \square\bar{T}=0 ,
\label{eq:N3}
\end{align}
where $h_\mu\equiv\partial^\nu h_{\mu\nu}$. Note also that Eqs.~(\ref{eq:N02}), (\ref{eq:N12}) and \eqref{eq:N22}
are already simplified by their preceding equations.

The $\eta^{\mu\nu}$-trace of Eq.~(\ref{eq:N02}) together with
Eq.~(\ref{eq:N01}) and $\partial^\mu b_\mu^{\T}=0$ yields
\begin{equation}
2\lambda= \partial^\nu h_\nu= \partial^\mu\partial^\nu h_{\mu\nu},
\label{eq:h=2lambda}
\end{equation}
implying that the multiplier field $\lambda$ imposing the unimodular constraint
equals half of the double divergence of $h_{\mu\nu}$. Taking the divergence
$\partial^\nu$ of Eq.~(\ref{eq:N02}) and using Eq.~\eqref{eq:h=2lambda}, we find
\begin{equation}
\square b_\mu^{\T} = - \partial_\mu\lambda,
\label{eq:Boxbmu}
\end{equation}
which, owing to the transversality of $b_\mu^{\T}$, also implies that $\lambda$ is the
massless simple-pole field
\begin{equation}
\square\lambda=0.
\label{eq:BoxLmd}
\end{equation}
The dual of Eq.~(\ref{eq:N03}) gives
\begin{equation}
\partial^{[\mu}h^{\nu]} = -\frac12 \varepsilon^{\mu\nu\rho\sigma}\partial_\rho B_\sigma,
\end{equation}
whose divergence $\partial_\nu$ yields, with the help of Eq.~(\ref{eq:h=2lambda}),
\begin{equation}
\square h_\mu= 2\partial_\mu\lambda.
\label{eq:BoxHmu}
\end{equation}
Acting d'Alembertian $\square$ on Eq.~(\ref{eq:N02}) and using
Eqs.~(\ref{eq:Boxbmu}), (\ref{eq:BoxLmd}) and (\ref{eq:BoxHmu}),
we obtain
\begin{equation}
\square^2 h_{\mu\nu}= 2\partial_\mu\partial_\nu\lambda.
\label{eq:Box2Hmunu}
\end{equation}
The EOM of $B^{\mu\nu}$ can be obtained from that of $b_\mu^{\T}$;
the dual of Eq.~(\ref{eq:bTBb}) at linearized level leads to
\begin{equation}
b_\mu^{\T} = \frac12 \varepsilon_{\mu\nu\rho\sigma}\partial^\nu B^{\rho\sigma} \ \rightarrow\
3\partial^{[\nu}B^{\rho\sigma]}= -\varepsilon^{\mu\nu\rho\sigma}b_\mu^{\T}.
\end{equation}
Taking the divergence $\partial_\nu$ of this equation and then
using the first relation $\partial_\nu B^{\nu\mu}=\partial^\mu B$ in Eq.~(\ref{eq:N04}),
we get a simple EOM for $B^{\mu\nu}$:
\begin{equation}
\square B^{\rho\sigma} = \varepsilon^{\rho\sigma\mu\nu}\partial_\mu b_\nu^{\T}\,.
\label{eq:BoxBmunu}
\end{equation}

Using Eq.~(\ref{eq:N04}) and the antisymmetry property of $B^{\mu\nu}$, and
taking the divergence $\partial^\rho$ of Eq.~(\ref{eq:N03}), we find
\begin{equation}
\square B=0,  \qquad  \square B_\mu=0.
\end{equation}
Similarly, from Eqs.~(\ref{eq:N11})--(\ref{eq:N13}),
we also find
\begin{equation}
\square C=\square\bar{C} =0, \qquad
\square C_\mu=\square\bar{C}^\mu=0\,.
\end{equation}
Equations~(\ref{eq:N21}) - (\ref{eq:N3}) also show that all
the fields with $|N_\FP| \geq2$ are of simple pole:
\begin{equation}
\square D = \square\bar{D}=
\square D_\mu= \square\bar{D}^\mu= \square T = \square\bar{T} = 0.
\end{equation}

\section{Identifying independent fields and BRST quartets}
\label{counting}

The present system contains several multipole fields up to tripole fields,
so that we generally have to decompose them into simple-pole modes in order
to count the independent particle modes in detail. For example, even for
the simplest dipole scalar field $\square^2\phi=0$, it is decomposed into two
simple-pole modes $\hat\phi$ and $\varphi:= \square\phi$, both satisfying simple-pole
EOMs $\square\hat\phi=0,\ \square\varphi=0$. Indeed, in terms of these
two simple-pole fields $\hat\phi$ and $\varphi$, the original dipole field
$\phi$ can be expressed as
\begin{equation}
\phi(x) = \hat\phi(x) - \calD_x \varphi(x),
\label{eq:Decomp}
\end{equation}
by using an integro-differential operator $\calD_x$ introduced by Nakanishi
and Lautrup\cite{Nakanishi:1966zz} long time ago:
\begin{equation}
\calD_x := \frac1{2\nabla^2}\left(x^0\partial_0 - 1/2 \right),
\end{equation}
which acts as an ``inverse" of $-\square$ in front of any simple-pole function
$f(x)$:
\begin{equation}
-\square\calD_x f(x)= f(x) \quad \text{if}\quad \square f(x)=0.
\end{equation}
To treat tripole fields, a similar operator $\calT_x$ would become necessary which acts as an ``inverse"
of $(-\square)^2$ in front of any simple-pole function as $(-\Box)^2 \calT_x f(x)= f(x)$.

However, we can avoid such an explicit but tedious procedure by adopting
the four-dimensional Fourier expansion\cite{Nakanishi:1966zz} of the fields defined by
\begin{equation}
\phi(x) = \frac1{\sqrt{(2\pi)^3}}\int d^4p\, \theta(p^0)\left[
\phi(p)e^{ipx} + \phi^\dagger(p)e^{-ipx} \right].
\end{equation}
Then, the four-dimensional operators $\phi(p)$ and $\phi^\dagger(p)$ annihilate and create
the multipole particles as they stand. The BRST singlet physical modes
must of course be simple-pole fields. Multipole fields are necessarily
unphysical and so will fall into BRST quartets, provided that we are treating
a consistent theory. We will see that this is actually the case in this UG
theory. We note that, when $\phi(x)$ is a simple-pole field, $\phi(p)$ is given
in terms of the usual annihilation operator $\phi(\bfp)$ by three-dimensional Fourier transform as
\begin{equation}
\phi(p) = \theta(p^0) \delta(p^2) \sqrt{2|\bfp|} \,\phi(\bfp).
\end{equation}

Let us now analyze independent four-dimensional Fourier modes for each ghost
number $N_\FP$ sector successively,
in the Lorentz frame in which the 3-momentum $\mbf{p}$ is along the $x^3$ axis:
\begin{equation}
p^\mu= \bigl(p^0, 0, 0, p^3\bigr), \quad \text{i.e.,}\quad
p^i=0\ (i=1,2), \
p^3=:|p|>0.
\label{eq:frame}
\end{equation}
In particular, in front of massless simple-pole fields $\phi(p)\propto\delta(p^2)$,
it reads
\begin{equation}
p^\mu\,\phi(p^2)=  \bigl( |p|, 0, 0, |p|  \bigr) \,\phi(p^2),
 \quad \text{i.e.,} \quad
p^0=p^3=|p|\ .
\label{eq:onshellp}
\end{equation}

For the task to identify all independent fields, it is easy and
transparent to consider the BRST transformation of the fields and to
identify the BRST quartets simultaneously. Actually we shall show that
all the independent fields other than the transverse graviton with helicity
$\pm2$ fall into BRST quartets which decouple from the physical sector (more
precisely, appear only in zero-norm combination in the physical subspace).
This means that all the other field components than those appearing explicitly as
members of the BRST quartets are all dependent fields or vanish.
In order to show this, we recall the BRST transformation of all the fields:
\begin{align}
\delta_\B \lambda&= 0,
\label{eq:BRSlambda}\\
\delta_\B h^{\mu\nu} &= -2\partial_{\phantom{T}}^{(\mu}c_{\T}^{\nu)}
= \partial^{(\mu}\varepsilon^{\nu)\tau\rho\sigma}\partial_\tau C_{\rho\sigma},
\label{eq:BRShmunu}\\
\delta_\B C_{\mu\nu} &= i (\partial_\mu D_\nu-\partial_\nu D_\mu),
\label{eq:BRSCmunu}\\
\delta_\B \bar C^{\mu\nu} &= i B^{\mu\nu}, \qquad  \delta_\B B^{\mu\nu}=0,
\label{eq:BRSbarCmunu}\\
\delta_\B \bar D^\mu&= \bar C^\mu, \hspace{.9em}\qquad  \delta_\B \bar C^\mu=0,
\label{eq:BRSbardmu}\\
\delta_\B B_\mu&=  C_\mu, \hspace{.9em}\qquad  \delta_\B C_\mu=0,
\label{eq:BRSBmu}\\
\delta_\B D_\mu&=  \partial_\mu T,\hspace{.8em}\qquad \d_\B T=0,
\label{eq:BRSdmu}\\
\delta_\B \bar T &=  i \bar D, \hspace{1.2em}\qquad  \delta_\B \bar D=0,
\label{eq:BRSbart}\\
\delta_\B C &=  iD, \hspace{1.2em}\qquad \delta_\B D=0,
\label{eq:BRSC}\\
\delta_\B \bar C &=  iB, \hspace{1em}\qquad  \delta_\B B =0 \,.
\label{eq:BRSbarC}
\end{align}

The BRST quartet is generally a pair of the BRST doublets which satisfies the
properties schematically drawn as
\begin{equation}
\xymatrix{
 A(p)\ \ar@{->}[r]^{\delta_{\text{B}}}\ar@{<.>}[rrd] 
                   & \ C(p) \ar@{<..>}[d]_{\text{inner-product}} &  \\
   &\ \bar C(p)\  \ar[r]_{\delta_{\text{B}}} & \ iB(p) ,
 }
\label{eq:BRSquartet}
\end{equation}
which means that
$ \big( A(p)\rightarrow C(p) \big)$ and $\big( \bar C(p)\rightarrow iB(p) \big)$ form a pair of
BRST doublets satisfying (assuming $A(p)$ a boson),
\begin{align}
\delta_{\B}  A(p) &= [iQ_\B,\,A(p)] = C(p), \nn
\delta_{\B}  \bar C(p) &= \{iQ_\B,\,\bar{C}(p) \} = iB(p),
\end{align}
and have nonvanishing inner product with each other:
\begin{equation}
\langle0| \bar{C}(p) C^\dagger(q) |0\rangle=
\langle0| \bar{C}(p)iQ_\B A^\dagger(q) |0\rangle= i\langle0| B(p) A^\dagger(q) |0\rangle
\propto\delta^4(p-q) \not=0,
\label{eq:Innerproduct1}
\end{equation}
or, equivalently, in terms of commutation relation,
\begin{align}
0&=\bigl\{ iQ_\B,\, [\bar{C}(p),\, A^\dagger(q)] \bigr\} =
[i B(p),\, A^\dagger(q)] - \{ \bar{C}(p),\, C^\dagger(q) \}  \nn
&\rightarrow\
[i B(p),\, A^\dagger(q)] = \{ \bar{C}(p),\, C^\dagger(q) \}
\propto\delta^4(p-q) \not=0.
\label{eq:Innerproduct2}
\end{align}
Let us denote this BRST quartet shown by the scheme (\ref{eq:BRSquartet})
simply as
\begin{equation}
\bigl( A(p) \rightarrow C(p);\ \bar C(p) \rightarrow iB(p) \bigr).
\label{eq:QuartetNotation}
\end{equation}

We now give the details of our analysis.

\subsection{$N_\FP=0$ sector}
\label{counting.1}

We begin with the fields with ghost number $N_\FP=0$. We have 10 component gravity $h_{\mu\nu}$ field,
1 unimodular multiplier field $\lambda$, plus 6 $B^{\mu\nu}$, 4 $B_\mu$
and 1 $B$ fields; thus, $10+1+6+4+1=22$ component fields in all.
Let us count/identify the independent fields among them, by using the EOMs~(\ref{eq:N01}) -- (\ref{eq:N05}).

We first note that the EOMs for the gravity field $h_{\mu\nu}$,
(\ref{eq:N01}) and (\ref{eq:N02}) exactly take the same form as those in the
GR theory in unimodular gauge, which we have presented in the previous paper~\cite{KNO}.
This holds provided that we identify the previous unimodular
gauge-fixing multiplier field (NL field) $b$ in GR with
the present unimodular multiplier field $\lambda$.
Moreover, the BRST transformations of $h^{\mu\nu}$ and $\lambda$ given in
Eqs.~(\ref{eq:BRSlambda}) and (\ref{eq:BRShmunu})
also take the same form as in the previous GR case.
In the previous GR case, the first equation (\ref{eq:BRSlambda})
holds because the NL multiplier
field $b$ identified with $\lambda$ here is the BRST daughter field of a FP
antighost called $\bar d$ there. The second equation (\ref{eq:BRShmunu})
also holds since it is merely the general coordinate transformation of the Einstein gravity theory.
An apparent difference is that $c_{\T}^\mu$ here is subject to the transversal
constraint $\partial_\mu c^\mu_{\T}=0$ (off-shell). However, the transversal
condition $\partial_\mu c^\mu=0$ for the FP ghost field also appeared as an
EOM in the previous GR case. We should note that only the
on-shell property is relevant here since we are analyzing the BRST
structure of the on-shell modes of asymptotic fields.

Therefore, from the previous result in the GR case, we immediately
see that we have the following 5 independent fields from 10 component
$h^{\mu\nu}(p)$ fields.
First of all, we have two BRST invariant simple-pole (hence physical) fields
\begin{align}
&h_{\T1}(p):= \half \bigl(h^{11} - h^{22}\bigr)(p),  \qquad
h_{\T2}(p):= h^{12}(p).
\end{align}
These transverse modes are BRST invariant because the transverse
momentum components $p^i\ (i=1,2)$ vanish by definition. Simple-pole
property $\square h_{\T j}(p)=0$ also follows from the EOM (\ref{eq:N02}) and $p^i=0$.
In addition to these two, we have a transverse vector field (hence possessing
3 independent components):
\begin{align}
&\chi^0(p) := \frac1{2p^0}\Bigl(h^{00}-\frac12\bigl(h^{11} +h^{22}\bigr)\Bigr)(p)
= \frac1{2p^0}\frac12\bigl(h^{00} + h^{33}\bigr)(p), \nn
&\chi^i(p) := \frac1{p^0}h^{0i}(p) \ \ (i=1,2), \nn
&\chi^3(p) := \frac1{2p^3}\Bigl(h^{33}-\frac12\bigl(h^{11}
+h^{22}\bigr)\Bigr)(p)
= \frac1{2p^3}\frac12\bigl(h^{00} + h^{33}\bigr)(p),
\label{eq:chimu}
\end{align}
satisfying transversality $p_\mu\chi^\mu(p)=p_0\chi^0(p)+p_3\chi^3(p)=0$.
So we can forget the redundant component $\chi^3(p)$ henceforth.
The second equality for the expression $\chi^0(p)$ (or $\chi^3(p)$) follows
from the tracelessness Eq.~(\ref{eq:N01}), $h:=\eta_{\mu\nu}h^{\mu\nu}=0$,
\begin{equation}
\left(h^{11}+h^{22}\right)(p) =
\left(h^{00}-h^{33}\right)(p).
\label{eq:h11+h22}
\end{equation}
This $\chi^\mu(p)$ field has a very simple BRST transformation property
\begin{equation}
\delta_\B \chi^\mu(p) = -i c^\mu_{\T}(p).
\label{eq:BRSchimu}
\end{equation}

If we rewrite $c^\mu_{\T}(p)$ in terms of unconstrained FP ghost fields
$C_{\mu\nu}(p)$, this BRST transformation law (\ref{eq:BRSchimu}) is written
more explicitly for the independent fields $\chi^0(p)$ and $\chi^i(p)$ as
\begin{align}
\bigl(\ \chi^0(p) \ \overset{\delta_\B}{\longrightarrow}\
& -ic^0_{\T}(p) = -p_3C_{12}(p)\ \bigr),
\label{eq:chi0doublet}\\
\bigl(\ \chi^i(p) \ \overset{\delta_\B}{\longrightarrow}\
& -i c^i_{\T}(p)
= \varepsilon^{ij}\left( p_3C_{0j}-p_0C_{3j}\right)(p)
= -\varepsilon^{ij}\frac1{p^3}\square C_{0j}(p)\ \bigr) \nn
&\hspace{9em} (i,j=1,2; \,
\varepsilon_{ij}=-\varepsilon_{ji},\, \varepsilon_{12}=+1 ),
\label{eq:chiidoublet}
\end{align}
where use has been made of the EOM $\partial^\mu C_{\mu i}(p)=0$ to rewrite
$C_{3i}(p)$ as $-(p^0/p^3)C_{0i}(p)$ in the last equality.

The partner BRST doublets which have nonvanishing inner products with these
three BRST doublets are now easily identified, respectively, as
\begin{align}
\delta_\B {\bar{C}}^{12}(p) &= i {B}^{12}(p),
\label{eq:barC12doublet}\\
\delta_\B {\bar{C}}^{0i}(p) &= i B^{0i}(p), \quad (i=1,2).
\label{eq:barC0idoublet}
\end{align}
For instance, the non-vanishing propagator $\VEV{\T C_{12}\,\bar{C}^{12}}\not=0$
means the non-vanishing innerproduct
$\langle0|C_{12}(p)\,\bar{C}^{12\dagger}(q) |0\rangle\not=0$.
The relevant innerproducts or commutation relations
can generally be read from the propagators, which we shall discuss
in detail in the next section.
If we use the quartet notation (\ref{eq:QuartetNotation}),
these three BRST quartets are denoted as
\begin{align}
&\bigl(\ \chi^0(p) \ \rightarrow\ -p^3C_{12}(p);\
\ \bar{C}^{12}(p) \ \rightarrow\ i {B}^{12}(p) \ \bigr),
\label{eq:FirstQuartet}\\
&\bigl(\ \varepsilon_{ij}\chi^j(p) \ \rightarrow\  (1/p^3)\square C_{0i}(p) ;\
\ {\bar{C}}^{0i}(p) \ \rightarrow\ i B^{0i}(p)\ \bigr),  \quad (i=1,2).
\label{eq:SecondQuartet}
\end{align}

Note that BRST quartets which have the graviton fields $h^{\mu\nu}$ as
their BRST parent components are {\it only these three} quartets.
This is because
the BRST transformation of $h^{\mu\nu}$ field is given as
Eq.~(\ref{eq:BRShmunu}) in terms of the transversal vector $c^\mu_\T$ so that
only three independent components of $h^{\mu\nu}$ can become BRST parents.

Thus aside from the two physical fields $h_{\T i}(p)$ there appear only three
components of $h^{\mu\nu}$;
$\chi^0(p)\propto(h^{00}+h^{33})(p)$ and $\chi^i(p)\propto h^{0i}(p)$ $(i=1,2)$ in these
BRST quartets. There still remain 5 components in $h^{\mu\nu}(p)$ which
have not yet appeared; they are given in suitable basis as
\begin{equation}
\bigl(h^{11}+h^{22}\bigr)(p),\quad
\bigl(h^{00}-h^{33}\bigr)(p),\quad
h^{3i}(p),\quad h^{03}(p).
\label{eq:dependentHmunu}
\end{equation}
In order for UG theory to have only two physical modes of transversal graviton, those 5 components
each must either vanish or become dependent field written in terms of those
independent fields $\chi^\mu(p)$ and/or independent components
of $B^{\mu\nu}$.
Three components of $B^{\mu\nu}$, $B^{12}(p)$ and $B^{0i}(p)$, appear in these
BRST quartets as the BRST daughter fields of the partner doublets
(\ref{eq:barC12doublet}) and (\ref{eq:barC0idoublet}), so they can be chosen
independent fields among 6 component $B^{\mu\nu}$. The rest 3 components
\begin{equation}
B^{03}(p), \quad B^{3i}(p),
\label{eq:dependentBmunu}
\end{equation}
must be dependent fields.

Let us now show successively by using Eqs.~(\ref{eq:N01}) -- (\ref{eq:N05})
that those 5 components of $h^{\mu\nu}(p)$ in Eq.~(\ref{eq:dependentHmunu})
and 3 components of $B^{\mu\nu}$ in Eq.~(\ref{eq:dependentBmunu}) are
dependent fields.

Start with the simple one to show that $B^{03}(p)$ and $B^{3i}(p)$ are
dependent fields.
The EOM (\ref{eq:N04}) gives a vector constraint
$\partial_\nu B^{\nu\mu}-\partial^\mu B=0$ with an index $\mu$. Since
$\partial_\nu B^{\nu\mu}$ is a transverse vector due to antisymmetry property of
$B^{\nu\mu}$, it gives only 3 constraints on $B^{\nu\mu}$ aside from implying
a simple pole field equation $\square B=0$ for $B$.
Explicitly, we obtain the following 3 constraints on $B^{\nu\mu}$:
\begin{align}
(\mu= 0\text{ or 3}):&\
p_3 B^{30}(p)  \text{ or } p_0B^{03}(p) = p^0B(p)= p^3 B(p) \ \rightarrow\
B^{03}(p) = -B(p), \nn
(\mu=i):&\ p_0 B^{0i}(p)+p_3 B^{3i}(p) =  0 \ \rightarrow\
B^{3i}(p)= -\frac{p_0}{p_3}B^{0i}(p) \quad (i=1,2).
\label{eq:B3iB0i}
\end{align}
Note that $p_0=-p^0$. This shows the desired dependent properties of the fields $B^{03}(p)$ and
$B^{0i}(p)$ provided that we also choose the scalar field $B(p)$ as an
independent field in addition to the BRST daughter $B^{0i}(p)$.
Later, we shall see that $B(p)$ also appears as BRST daughter field
of another BRST quartet.

We now show the dependency of the $h^{\mu\nu}(p)$ components
in Eq.~(\ref{eq:dependentHmunu}).
First, the trace component $h:=\eta_{\mu\nu}h^{\mu\nu}=0$ vanishes by
the unimodularity Eq.~(\ref{eq:N01}) as noted above. This implies
that the variable $h^{11}+h^{22}$ becomes dependent
field written in terms of $h^{00}-h^{33}$ as written in (\ref{eq:h11+h22}).
The latter field $h^{00}-h^{33}$ will be shown, in Eq.~(\ref{eq:H00-33})
below, equal to the $B^{12}$ field which is the BRST daughter field.

Next, recall that our gauge-fixing by transverse vector multiplier
field $b_\mu^{\T}$, yields the EOM (\ref{eq:N03}) as gauge-fixing conditions
of transverse de Donder gauge.
Apparently Eq.~(\ref{eq:N03}) possesses 6 components, but it
actually implies only the following 3 independent constraints
on $h^\mu(p)$:\footnote{
The constraint equation (\ref{eq:N03}) for $\rho=0, \sigma=3$ yields
$p_0B_3(p)-p_3B_0(p)=0$, which is identical with $\partial^\mu B_\mu(p)=0$ owing to
the on-shell momentum equality (\ref{eq:onshellp}) on the simple-pole field $B_\mu(p)$.}
Explicitly, they read
\begin{align}
\text{($\rho=0$ or 3, $\sigma=i$)} &:~  h^i(p) = \varepsilon^{ij} B_j(p),\quad
(i,j = 1,2,\ \varepsilon^{ij}=-\varepsilon^{ji}),
\label{eq:Hi}\\
\text{($\rho=1$, $\sigma=2$)} &:~  p^3h^0(p) - p^0h^3(p) = 0.
\label{eq:H0H3}
\end{align}
Here in the first equation, we have already set $p^0=p^3=|p|$ in front of
the simple-pole fields $B_\mu(p)$ and $h^i(p)$ and factored out $|p|$.
(But we cannot do so for the second equation because
$h^0(p)$ and $h^3(p)$ are not simple pole but dipole fields.)
If we express $h^\mu$ in terms of $h^{\mu\nu}$ by $h^\mu(p)=ip_\nu h^{\mu\nu}(p)$,
these constraint Eqs.~(\ref{eq:Hi}) and (\ref{eq:H0H3}) are rewritten as
\begin{align}
&p_0h^{0i}(p) +p^3 h^{3i}(p) = -i\varepsilon^{ij}B_j(p), \ (i,j=1,2),
\label{eq:Hi2} \\
&-p^3p^0 ( h^{00}  + h^{33} )(p) + (p_0^2+p_3^2) h^{03}(p) =0.
\end{align}
Therefore, owing to the gauge condition (\ref{eq:N03}), three
components $h^{3i}(p)$ and $h^{03}(p)$ now become dependent fields:
\begin{align}
h^{3i}(p) &= \frac1{p_3}\bigl(p_0^2\chi^i(p) -i\varepsilon^{ij}B_j(p)\bigr), \ (i,j=1,2),
\label{eq:H3i} \\
h^{03}(p) &= \frac{4p_0^2p^3}{p_0^2+p_3^2}\chi^0(p),
\label{eq:H03}
\end{align}
where use has been made of the relations $h^{0i}(p)=p^0\chi^i(p), \
\bigl(h^{00}+h^{33}\bigr)(p)=4p^0\chi^0(p)$ in Eq.~(\ref{eq:chimu}).
$B_i(p)$ on the RHS will be shown below to be a dependent field given
in terms of $\chi^i$ and $B^{0i}$.

Now that we have shown $1+3$ fields, $h^{11}+h^{22}$, $h^{3i}$ and $h^{03}$,
to be dependent variables, we need one more constraint, an equation giving
$h^{00}-h^{33}$ in terms of $B^{12}$ field as announced before.
It can be obtained from the EOMs (\ref{eq:N02}) with indices
$(\mu\nu)=(0,0)$ and (3,3),
\begin{align}
&\frac12(p_0^2-p_3^2) h^{00} + \bigl( -p_0^2h^{00} + p^0p_3 h^{03} \bigr)
-ip^0b^0_{\T} -\frac12\lambda=0 , \nn
&\frac12(p_0^2-p_3^2) h^{33} + \bigl( +p^3p_0h^{03} + p_3^2 h^{33} \bigr)
-ip^3b^3_{\T} + \frac12\lambda=0.
\end{align}
Adding them and using the transversality of $b_{\T}^\mu$, $p_0b^0_{\T}+p_3b_{\T}^3=0$,
we find a constraint
\begin{align}
&-\frac12(p_0^2+p_3^2) \left(h^{00}-h^{33}\right)(p) -2ip^0b^0_{\T}(p) =0.
\end{align}
Since 
$b^\mu_{\T} := \half\varepsilon^{\mu\nu\rho\sigma}\partial_\nu B_{\rho\sigma}$ implies
$b^0_{\T}(p)=ip^3 B^{12}(p)$, this gives the desired
expression for $h^{00}-h^{33}$ in terms of $B^{12}$:
\begin{equation}
\bigl(h^{00}-h^{33}\bigr)(p) =
\frac{4p^0p^3}{p_0^2+p_3^2}B^{12}(p).
\label{eq:H00-33}
\end{equation}

That is all for the 10 components, $h^{\mu\nu}$.
As other fields in this $N_\FP=0$ sector, we still have unimodular
multiplier field $\lambda$ and a vector field $B_\mu$.

The field $\lambda$ is clearly a dependent
field which can be expressed in many ways, as already derived in Eqs.~(\ref{eq:h=2lambda}), (\ref{eq:Boxbmu}),
(\ref{eq:BoxHmu}) and (\ref{eq:Box2Hmunu}) from the EOMs (\ref{eq:N02}) and (\ref{eq:N03}).
The most remarkable and important expression among them is
Eq.~(\ref{eq:Boxbmu}), the $\mu=0$ component of which, in particular, gives
\begin{equation}
\square b_0^{\T}(p) = ip^0 \lambda(p).
\label{eq:Boxb0}
\end{equation}
Substituting $b_0^{\T}(p)=-ip^3B^{12}(p)$ and dividing both sides by
$ip^3=ip^0$ valid on simple-pole fields $\lambda$ and $\square B^{12}$, we find
\begin{equation}
\lambda(p) = -\square B^{12}(p).
\label{eq:lambda=B12}
\end{equation}
This equation (\ref{eq:lambda=B12}), or (\ref{eq:Boxb0}), says that
the unimodular multiplier field $\lambda$ in the UG theory
in fact becomes identical with the gauge-fixing multiplier $B^{12}$
field (NL field) as if the unimodular condition were imposed as a gauge-fixing
condition just like in GR theory in unimodular gauge.
This is a key equation which makes the quantum UG theory consistent.

Finally in this subsection, we discuss which components are independent
in the vector field $B_\mu$. We have the transversality EOM (\ref{eq:N05}):
\begin{equation}
p^0B_0(p) +p^3B_3(p) = 0 \ \rightarrow\ B_3(p) = - B_0(p),
\end{equation}
so that we regard $B_3(p)$ as a dependent field expressible in terms of
$B_0(p)$. We shall see that $B_0(p)$ becomes a BRST parent field of a
BRST quartet appearing at the next step.
The other transverse components $B_i(p)$ $(i=1,2)$, at first sight, look like
independent fields, but actually turn out to be dependent fields
written in terms of $\chi^i(p)$ and $B^{0i}(p)$, as announced below
Eq.~(\ref{eq:H3i}). This comes from the gravity EOM (\ref{eq:N02})
with index $\mu=0, \ \nu=i$, which reads
\begin{equation}
\square h^{0i}(p) -ip^0 \left(h^i(p)+\eta^{ij}b_j^{\T}(p)\right) =0.
\end{equation}
Since $h^i(p)$ here is $\varepsilon^{ij}B_j(p)$ by Eq.~(\ref{eq:Hi}),
this gives the desired expression for $B_i(p)$:
\begin{align}
B_i(p) &= \varepsilon_{ij} \left( \eta^{jk}b_k^{\T}(p) +i \frac1{p^0}\square h^{0j}(p)\right)\nn
&= i\square\left( -\frac1{p^0}\eta_{ij}B^{0j}(p) +\varepsilon_{ij}\chi^j(p)\right).
\end{align}
Here in going to the second line, we have rewritten
$b_j^{\T}(p)$ in terms of $B^{0i}(p)$ which follows
from its definition and $B^{3i}(p)=-(p_0/p^3)B^{0i}(p)$
in Eq.~(\ref{eq:B3iB0i}) as
\begin{align}
b_j^{\T}(p)
&= -\half \varepsilon_{j\nu\rho\sigma}\partial^\nu B^{\rho\sigma}(p) 
= - \varepsilon_{ij}\bigl(ip^0B^{3i}(p)-ip^3B^{0i}(p)\bigr) \nn
&= i\varepsilon_{ij}\frac1{p^3}(p_0^2 -p_3^2) B^{0i}(p)
= i\varepsilon_{ij}\frac1{p^0}\square B^{0i}(p)\,, 
\end{align}
where $p^3$ is replaced by $p^0$ in the last step in front of
the simple-pole field $\square B^{0i}(p)$.

The independent fields in this $N_\FP=0$ sector identified in
this subsection are summarized in the first line of Table 1;
$h_{\T1}$ and$h_{\T2}$ are transversal physical graviton with helicity $j=\pm2$,
$\chi^0$ and $\chi^i$ are BRST parents, while $B^{12}$ and $B^{0i}$ are
BRST daughters, of the first step BRST quartets (\ref{eq:FirstQuartet})
and (\ref{eq:SecondQuartet}). The rest fields, scalar $B$ and time component
$B_0$ of the vector field $B_\mu$, will appear in the BRST quartets in the
next step. (This should be obvious since $B_\mu$ and $B$ are the
member fields of the multiplier doublets in the second and third
steps of gauge fixing, respectively.)

\begin{table}[htb]
\caption{List of independent fields. $i$ denotes transverse directions 1 and 2.}
\label{table:1}
\begin{center}
\begin{tabular}{|r|ccc|} \hline\hline
$N_\FP=0$
 & \Tspan{$ 
   h_{\T1},\ h_{\T2};\quad \chi^0,\ \chi^i $}
 &;& \Tspan{$ 
   {B}^{0i},\ {B}^{12},\ B,\ B_0 $} \\ \hline
$|N_\FP|=1$    & \Tspan{$ 
   {C}_{0i},\ {C}_{12},\ C,\ C_0 $}
  &;& \Tspan{$ \bar{C}^{0i},\ \bar{C}^{12},\ \bar{C},\ \bar{C}^0$} \\ \hline
$|N_\FP|=2$    & \Tspan{$ D_0,\ D_i,\ D $}
  &;&  \Tspan{$ \bar{D}^0,\ \bar{D}^i,\ \bar{D} $} \\ \hline
$|N_\FP|=3$      & $T$ &;& \Tspan{$\bar{T}$}              \\ \hline
\end{tabular}
\end{center}
\end{table}

\subsection{$N_\FP\not=0$ sector}
\label{counting.2}

Now all the fields with $N_\FP\not=0$ are ghosts which should decouple
from the physical sector, so that all the independent fields belong
to BRST quartets in some steps of gauge fixing.

Consider the fields with $N_\FP=\pm1$, $C_{\mu\nu}, C_\mu$ and
$\bar{C}^{\mu\nu}, \bar{C}^\mu$.
First begin with the antighost part.
Among 6 components of $\bar{C}^{\mu\nu}$, $\bar{C}^{12}$ and
$\bar{C}^{0i}\ (i=1,2)$ are independent fields which already appeared
in the first step BRST quartets,
(\ref{eq:FirstQuartet}) and
(\ref{eq:SecondQuartet}). The other 3 components,
$\bar{C}^{03}$ and $\bar{C}^{3i}$,
must be dependent fields. Indeed, we can show this by the second equation
in the EOM (\ref{eq:N11}):
\begin{equation}
\partial_\nu\bar{C}^{\nu\mu} -\partial^\mu\bar{C}=0,
\label{eq:N11b}
\end{equation}
which is the gauge condition on $\bar{C}^{\mu\nu}$ imposed by the second step
gauge-fixing with transversal vector multiplier $C_\mu$. This takes exactly
the same form as the previous Eq.~(\ref{eq:N04}),
$\partial_\nu B^{\nu\mu} -\partial^{\mu}B=0$,
for $B^{\mu\nu}$.
[This coincidence is actually a result of BRST invariance; since
$\delta_\B\bar{C}^{\mu\nu}=iB^{\mu\nu}, \ \delta_\B\bar{C}=iB$.
Equation~(\ref{eq:N04}) is merely the BRST transform of
this equation (\ref{eq:N11b}).]
Therefore, from the previous result (\ref{eq:B3iB0i}) for $B^{\mu\nu}$,
we immediately obtain
\begin{equation}
\bar{C}^{03}= -\bar{C}(p), \qquad \bar{C}^{3i}=-\frac{p_0}{p_3}\bar{C}^{0i}(p)
\quad (i=1,2),
\label{eq:barC03-3i}
\end{equation}
showing that $\bar{C}^{03}$ and $\bar{C}^{3i}$ are dependent fields, if
$\bar{C}(p)$ is chosen as an independent field in addition to the BRST
parents $\bar{C}^{0i}$. We shall see below that $\bar{C}(p)$ is the
BRST parent of the daughter $B(p)$ in a BRST quartet.

From the vector $\bar{C}^\mu$, we can show that only the $\bar{C}^0$ is
independent, similarly to the previous $B^\mu$ field.
$\bar{C}^3(p)$ is dependent because of Eq.~(\ref{eq:N13}),
$p_0\bar{C}^0+p_3\bar{C}^3=0$. For the transverse components
$\bar{C}^i$, we have the EOM (\ref{eq:N12})
whose $\mu=0,\ \nu=i$ components tell us that they are dependent on $\bar{C}^{0i}$:
\begin{equation}
\square^2\bar{C}^{0i}(p)+ip^0\bar{C}^i(p)=0 \ \rightarrow\
\bar{C}^i(p) = \frac{i}{p^0}\square^2\bar{C}^{0i}(p).
\end{equation}

Next consider the FP ghost part, $C_{\mu\nu}$ and $C_\mu$.
The situation is almost parallel to the antighost part,
because the EOM is invariant under the FP ghost conjugation
\begin{eqnarray}
&&C_{\mu\nu} \leftrightarrow \bar C^{\mu\nu}, \qquad
C_{\mu} \leftrightarrow  \bar C^{\mu}, \qquad
C \leftrightarrow -\bar C, \\
&&
D_{\mu} \leftrightarrow \bar D^{\mu}, \qquad
D \leftrightarrow \bar D, \qquad
T \leftrightarrow \bar T,
\end{eqnarray}
although the BRST transformation is not.
Among 6 components of $C_{\mu\nu}$, $C_{12}$ and $\square C_{0i}\ (i=1,2)$
already appeared in the first step BRST quartets, (\ref{eq:FirstQuartet})
and (\ref{eq:SecondQuartet}). Note here that $\square C_{0i}$ with d'Alembertian
$\square$ is appearing contrary to the $\bar{C}^{0i}$ in the above antighost case.
The d'Alembertian operator $\square$ acting on $C_{0i}$ projects out the
simple-pole part contained in $C_{0i}$, which we denote as $\hat{C}_{0i}$
with hat $\hat{\phantom{a}}$ symbol in distinction from the whole $C_{0i}$.
So in this FP ghost sector, the simple pole part $\hat{C}_{0i}$ is not
contained in the first-step BRST quartets, but become BRST parents in the
second-step BRST quartets; indeed, since $D_\mu$ is of simple pole,
the BRST transformation law (\ref{eq:BRSCmunu}) gives
\begin{equation}
\delta_\B \hat{C}_{0i}(p) = -p_0 D_i(p).
\label{eq:BRSbarC0i}
\end{equation}
Thus the whole parts (simple-pole and dipole or higher-pole parts) of
$C_{0i}$ are seen to be independent fields.

The other 3 components of $C_{\mu\nu}$, $C_{03}$ and $C_{3i}$, are
dependent fields, provided that the  scalar $C$ is chosen as
an independent field.
This is clear since the FP-conjugation
invariance of the EOM guarantees that Eq.~(\ref{eq:barC03-3i})
with $\bar{C}^{\mu\nu}$ there replaced by $C_{\mu\nu}$ holds.
Similarly, only the $C_0$ component is independent among the vector $C_\mu$.
Thus we have shown that the independent fields in $N_\FP=\pm1$ sector
are as given in the second line in Table 1.

Now at this stage, the remaining independent fields with $N_\FP=\pm1$,
which can become BRST parents of doublets but have not appeared in the
previous step BRST quartets, are $\hat{C}_{0i},\ \bar{C}$ and $C$.
We have already given the BRST doublet for $\hat{C}_{0i}$ in
Eq.~(\ref{eq:BRSbarC0i}). The BRST doublets for $\bar{C}$ and $C$
are given by (\ref{eq:BRSbarC}) and (\ref{eq:BRSC}),
\begin{align}
\delta_\B \bar{C}(p)&= i B(p), \label{eq:BRSbarCp}\\
\delta_\B C(p)&= i D(p). \label{eq:BRSCp}
\end{align}
The partner BRST doublets which have non-vanishing innerproducts with
these BRST doublets, (\ref{eq:BRSbarC0i})--(\ref{eq:BRSCp}), are found, respectively, as
\begin{align}
\delta_\B \bar D^{\,i}(p) &= \bar{C}^i(p)=\frac{i}{p^0}\square^2\bar{C}^{0i}(p),
\label{eq:BRSbarDi} \\
\delta_\B B_0(p) &= C_0(p), \label{eq:BRSB0} \\
\delta_\B \bar{D}^{\,0}(p) &= \bar{C}^0(p). \label{eq:BRSbarD0}
\end{align}
Thus, at this stage, we have the following three BRST quartets:
\begin{align}
&\bigl(\
\bar{D}^{\,i}(p) \ \rightarrow\ \bar{C}^i(p); \ \  
\hat{C}_{0i}(p) \ \rightarrow\ -p_0 D_i(p)
\ \bigr),
\quad (i=1,2)
\label{eq:FirstQuartet2}\\
&\bigl(\
B_0(p)\ \rightarrow\ C_0(p);\ \
\bar{C}(p) \ \rightarrow\ iB(p)
\ \bigr),
\label{eq:SecondQuartet2} \\
&\bigl(\
\bar{D}^{\,0}(p) \ \rightarrow\ \bar{C}^0(p);\ \
C(p) \ \rightarrow\ i D(p)
\ \bigr),
\label{eq:ThirdQuartet2}
\end{align}
where we have put first the BRST doublets which have boson parents
inside the quartets according to our convention in (\ref{eq:QuartetNotation}).
In the quartet (\ref{eq:FirstQuartet2}), we have written
$\bar{C}^i(p)$ in place of $(i/p^0)\square^2\bar{C}^{0i}(p)$, for notational simplicity.

We now see that we have almost finished;
all the remaining independent fields with $|N_\FP|\leq1$ listed in Table 1,
$B,\ B^0,\ \hat C_{0i},\ C,\ C_0,\ \bar C,\ \bar C^0$ and $\square^2\bar{C}^{0i}$,
which are not contained in the previous BRST quartets
(\ref{eq:FirstQuartet}) and (\ref{eq:SecondQuartet}),
have appeared as members
in these three BRST quartets (\ref{eq:FirstQuartet2})--(\ref{eq:ThirdQuartet2}).

Let us finish our task of this section by considering the
ghost fields with $|N_\FP|=2$ and 3, $D_\mu, \ D,\ \bar{D}^\mu,\ \bar D$
and $T,\ \bar T$.
Already the components $D_i, D$ and $\bar D^i,\ \bar D^0$ appeared
in these three BRST quartets.
From the EOMs (\ref{eq:N21}) -- (\ref{eq:N3}), we can take the components
$D_0,\ D_i,\ D,\ T$ and
$\bar D^0,\ \bar D^i,\ \bar D,\ \bar T$ as independent fields
as listed in Table 1. Therefore, the remaining independent fields
are only the four components $D_0,\ \bar D,\ T,\bar T$. They all appear
in the pair of BRST doublets
\begin{align}
\delta_\B D_0(p) &= ip_0 T(p), \nn
\delta_\B \bar T(p) &= i\bar D(p),
\end{align}
forming the last BRST quartet:
\begin{equation}
\bigl(\ D_0(p) \ \rightarrow\ ip_0T(p) ;\ \
\bar T(p) \ \rightarrow\ i\bar D(p) \bigr).
\label{eq:LastQuartet}
\end{equation}

We have thus finished the proof that all the independent fields
other than the physical transverse graviton modes $h_{\T i}(p)$ fall
into the BRST quartets given in Eqs.~(\ref{eq:FirstQuartet}),
(\ref{eq:SecondQuartet}), (\ref{eq:FirstQuartet2})--(\ref{eq:ThirdQuartet2})
and (\ref{eq:LastQuartet}).

\section{Metric structure of BRST quartets}
\label{metric}

For any free field, once the propagator is found, its spectral function is
determined and hence any two point functions can be found. So, in particular,
we can find the commutation relations (CRs) of creation/annihilation
operators by 4D Fourier expansion directly from the form of the propagators.
The translation rule from the propagator to CR is
\begin{equation}
\begin{array}{ccc}
\text{propagator}\ \langle\phi_i\, \phi_j \rangle&&  \text{CR}\  [\phi_i(p),\, \phi^\dagger_j(q)]
\\[.1ex]
\hline
\phantom{\bigg|}
\displaystyle \frac1{i} \Bigl[ \frac1{p^2},\  \frac1{p^4},\
\frac1{p^6} \Bigr] 
&\leftrightarrow&
\Bigl[ \delta(p^2),\  -\delta'(p^2),\ \frac12\delta''(p^2) \Bigr] \theta(p^0)\delta^4(p-q)
\end{array}.
\label{eq:translation}
\end{equation}
So, for instance, in free QED with gauge parameter $\alpha$,
we have the photon propagator
\begin{equation}
\langle A_\mu A_\nu\rangle= \frac1i \, \frac{\eta_{\mu\nu}-(1-\alpha)p_\mu p_\nu/p^2}{p^2},
\label{eq:AbbrevProp}
\end{equation}
from which we can immediately find the following CR:
\begin{equation}
[A_\mu(p),\, A_\nu^\dagger(q)]
= \left(\eta_{\mu\nu} \delta(p^2) + (1-\alpha)p_\mu p_\nu\delta'(p^2)\right)
\theta(p^0)\delta^4(p-q).
\end{equation}
Here in Eq.~(\ref{eq:AbbrevProp}) and henceforth we use an abbreviated notation
for the propagator in momentum space:
\begin{equation}
\langle A\,B \rangle := \int d^4x\, e^{-ipx} \VEV{ \T A(x)\,B(0) }\,.
\end{equation}

Now, let us confirm the BRST quartet CRs of the form (\ref{eq:Innerproduct2})
explicitly for the first two
quartets (\ref{eq:FirstQuartet}) and (\ref{eq:SecondQuartet});
for those two quartets, we have to compute the commutators
\begin{align}
[ i{B}^{12}(p),\, \chi^{0\,\dagger}(q) ]
&=
\{ \bar{C}^{12}(p),\, -q^3{C}_{12}^\dagger(q) \},
\label{eq:quartetCR1}\\
{}[ iB^{0i}(p),\, \varepsilon_{jk}\chi^{k\,\dagger}(q) ] &=
\{ \bar{C}^{0i}(p),\, \frac1{q^3}\square C_{0i}^\dagger(q) \} \quad
(\text{no sum over $i$}).
\label{eq:quartetCR2}
\end{align}
We already computed all the propagators which are generally given by
$i$ times the inverse two point vertices, $i\times{\Gamma^{(2)}}^{-1}$,
so the relevant four
propagators can be read from Eqs.~(\ref{eq:GammaInNF0}) and
(\ref{eq:GammaInNF1}) as
\begin{align}
&\EV{ B^{12}\ (h^{00}+h^{33}) }
= i \frac2{-(p^2)^2}\left(\varepsilon^{12\lambda0}p^0p_\lambda
+\varepsilon^{12\lambda3}p^3p_\lambda\right)
= 4i \frac1{(p^2)^2}p^0p_3,
\label{eq:B12-Hprop} \\
&\EV{ B^{0i}\ h^{0j} }
= i \frac1{-(p^2)^2}\varepsilon^{0i3j}p^0p_3
= i \frac1{(p^2)^2}\varepsilon^{ij}p^0p_3,
\label{eq:B0i-Hprop} \\
&\EV{ \bar C^{12}\ C_{12} }=  \frac1{(-p^2)^2},
\label{eq:C12prop}\\
&\EV{ \bar C^{0i}\ C_{0j} }=
\frac1{(-p^2)^2}\left( \delta^i_j - \frac1{p^2}p^0p_0\delta^i_j \right)
=
\delta^i_j\frac{p_3^2}{(p^2)^3}.
\label{eq:C0iprop}
\end{align}
Using the translation rule (\ref{eq:translation}), we can find, for instance,
from Eq.~(\ref{eq:B12-Hprop}) the commutator
\begin{equation}
\bigl[B^{12}(p),\ (h^{00}+h^{33})^\dagger(q)\bigr]= 4p^0p_3\delta'(p^2)\theta(p^0)\delta^4(p-q),
\end{equation}
so that
\begin{align}
&\bigl[iB^{12}(p),\ \chi^{0\dagger}(q)\bigr]
=\frac{i}{4q^0}\bigl[B^{12}(p),\ (h^{00}+h^{33})^\dagger(q)\bigr]
=ip_3\delta'(p^2)\theta(p^0)\delta^4(p-q).
\label{eq:B12-HCR}
\end{align}
In the same way we obtain from Eqs.~(\ref{eq:B0i-Hprop}) -- (\ref{eq:C0iprop})
\begin{align}
&\bigl[iB^{0i}(p),\ \varepsilon_{jk}\chi^{k\dagger}(q)\bigr]
=\frac{i}{q^0}\varepsilon_{jk}\bigl[B^{0i}(p),\ h^{0k\dagger}(q)\bigr]
=i\delta^i_j p_3\delta'(p^2)\theta(p^0)\delta^4(p-q),
\label{eq:B0i-HCR} \\
&\bigl\{ C^{12}(p),\ -q^3C_{12}^\dagger(q) \bigr\}
= iq^3\delta'(p^2)\theta(p^0)\delta^4(p-q),
\label{eq:C12_CR}\\
& \bigl\{ \bar C^{0i}(p),\ \frac1{q^3}\square C_{0j}^\dagger(q) \bigr\}
= \frac1{q^3}(-q^2) i \delta^i_j p_3^2 \frac12 \delta''(p^2)
= i \delta^i_j p^3 \delta'(p^2) \theta(p^0)\delta^4(p-q).
\label{eq:C0i_CR}
\end{align}
These confirm non-vanishing (anti-)commutation relations
(\ref{eq:quartetCR1}) and (\ref{eq:quartetCR2}) for the
first and second BRST quartets, respectively, as
\begin{align}
[ i{B}^{12}(p),\, \chi^{0\,\dagger}(q) ]
&=
\{ \bar{C}^{12}(p),\, -q^3{C}_{12}^\dagger(q) \}
=iq^3\delta'(p^2)\theta(p^0)\delta^4(p-q),
\label{eq:quartetCR1A}\\
{}[ iB^{0i}(p),\, \varepsilon_{jk}\chi^{k\,\dagger}(q) ] &=
\{ \bar{C}^{0i}(p),\, \frac1{q^3}\square C_{0i}^\dagger(q) \}
=
i \delta^i_j p^3 \delta'(p^2) \theta(p^0)\delta^4(p-q).
\label{eq:quartetCR2A}
\end{align}

We also note that these are dipole commutation relations $\propto\delta'(p^2)$.
So, if we decompose these fields into simple-pole fields, each of these two
BRST quartets in fact represents a pair of BRST quartets; more explicitly,
consider the first BRST quartet (\ref{eq:FirstQuartet}). Then, acting
d'Alembertian $\square$ of $p$ or $q$ on CRs
(\ref{eq:quartetCR1A})
and using $\square_p\delta'(p^2)=(-p^2)\delta'(p^2)=\delta(p^2)$, we have
\begin{align}
[ i\square{B}^{12}(p),\, \chi^{0\,\dagger}(q) ]
&=\{ \square\bar{C}^{12}(p),\, -q^3{C}_{12}^\dagger(q) \}
=iq^3\delta(p^2)\theta(p^0)\delta^4(p-q),
\label{eq:quartetCR1F}\\
[ i{B}^{12}(p),\, \square\chi^{0\,\dagger}(q) ]
&=\{ \bar{C}^{12}(p),\, -q^3\square{C}_{12}^\dagger(q) \}
=iq^3\delta(p^2)\theta(p^0)\delta^4(p-q).
\label{eq:quartetCR1R}
\end{align}
Note that $\phi(x)=B^{12},\ C_{12}, \bar C^{12}$ are dipole fields
satisfying $\square^2\phi=0$. Although $\chi^0$ is a tripole field, we can treat
it as if it were a dipole field in this BRST quartet since
its tripole part is given by $\lambda(x)$ as is seen in Eq.~(\ref{eq:BoxHmu})
which is BRST invariant and commutative with $B^{12}$ and hence can
contribute to neither $\delta_\B\chi^0$ nor $[\chi^0(p),\ B^{12}(q)]$.
As explained in Eq.~(\ref{eq:Decomp}), dipole field $\phi(x)$ generally
have two simple-pole modes; the genuine simple-pole mode $\hat\phi$ and
dipole-part $\square\phi$.

The Eqs.~(\ref{eq:quartetCR1F}) and (\ref{eq:quartetCR1R}) mean
that non-vanishing (anti-)CRs exist between
the dipole part $\square B^{12}$ of $B^{12}$ and genuine simple-pole
part $\hat\chi^0$ of $\chi^0$, and between
the dipole part $\square\bar{C}^{12}$ of $\bar{C}^{12}$ and genuine simple-pole
part $\hat{C}_{12}$ of $C_{12}$.
So these Eqs.~(\ref{eq:quartetCR1F}) and (\ref{eq:quartetCR1R}) imply the Ward-Takahashi identity
(\ref{eq:Innerproduct2}) for the following two BRST quartets of simple-pole fields, respectively:
\begin{align}
&\bigl(\ \hat\chi^0(p) \ \rightarrow\ -p^3\hat{C}_{12}(p);\
\ \square\bar{C}^{12}(p) \ \rightarrow\ i \square{B}^{12}(p) \ \bigr),
\label{eq:FirstQuartetR}\\
&\bigl(\ \square\chi^0(p) \ \rightarrow\ -p^3\square C_{12}(p);\
\ \hat{\bar{C}}^{12}(p) \ \rightarrow\ i \hat{B}^{12}(p) \ \bigr).
\label{eq:FirstQuartetF}
\end{align}
It should be noted here that the Lagrange multiplier field $\lambda$ was identified with
the $\square B^{12}$ as shown in \eqref{eq:lambda=B12}, and
the first equation~\eqref{eq:FirstQuartetR} shows that it makes a BRST quartet.
Thus we have the ghost and antighost associated with the unimodular condition,
and this makes the counting of dof correct.

In the same way, the second BRST quartet is seen to represent the following
two BRST quartets of simple-pole fields
\begin{align}
&\bigl(\ \varepsilon_{ij}\hat\chi^j(p) \ \rightarrow\  (1/p^3)\widehat{\square C}_{0i}(p) ;\
\ \square{\bar{C}}^{0i}(p) \ \rightarrow\ i \square B^{0i}(p)\ \bigr)  \quad (i=1,2).
\label{eq:SecondQuartetR} \\
&\bigl(\ \varepsilon_{ij}\square\chi^j(p) \ \rightarrow\  (1/p^3)\square^2C_{0i}(p) ;\
\ \hat{\bar{C}}^{0i}(p) \ \rightarrow\ i \hat{B}^{0i}(p)\ \bigr)  \quad (i=1,2).
\label{eq:SecondQuartetF}
\end{align}
where $\widehat{\square C}_{0i}$ denotes the simple-pole part of the dipole field $\square C_{0i}$.

For the other BRST quartets (\ref{eq:FirstQuartet2})--(\ref{eq:ThirdQuartet2}) and (\ref{eq:LastQuartet}),
all their members are of simple pole, and the confirmation of the CRs is
much easier, and we omit these.

\section{Discussions}
\label{discussions}

In this paper we have formulated covariant BRST quantization of UG by gauge fixing only TDiff.
We have achieved this using antisymmetric tensors for the reparametrization ghosts which automatically
satisfy transverse condition. It turned out that the kinetic terms for the ghosts and antighosts have
new gauge invariance which must be gauge fixed. This is the well-known phenomenon as ghosts for
ghosts~\cite{Townsend:1979hd,Kimura:1980zd,HKO}.
We then gauge fixed the invariance, and then this requires further gauge fixing.
We have succeeded in fixing all the gauge invariances, which was confirmed by the existence of the propagators
for all fields. We have classified how many independent modes exist, and have shown that most of these modes
cancel out, leaving only two dofs corresponding to the two transverse modes of spin 2 graviton.
Our key observation is that the original Lagrange multiplier field $\la$ becomes a BRST daughter
and there exist associated ghost and antighost, making the counting of dof correct.
Even though we have many ghosts, we were able to make covariant quantization without using nonlocal projectors
and the origin of the ghosts is now clearly identified. In this sense the formulation is transparent.

In our previous paper~\cite{KNO}, we made BRST quantization of GR in the unimodular gauge in order to cast
light on the covariant quantization of UG. We tried to relate the resulting gauge-fixed theory to UG
by making Fourier transform with respect to the cosmological constant. However the attempt was not quite
successful. There are two problems in that formulation.

(1) In the BRST quantization, physical states are characterized by a subsidiary condition~\cite{KO1977}.
This condition requires that the vacuum expectation value (VEV) of all the BRST daughter fields should vanish on
the physical states. In GR in the unimodular gauge, the multiplier field $\la'$ in \p{eq:UGaction1} is precisely
such a field imposing the gauge condition, and its VEV must vanish. This leaves the nonvanishing
cosmological constant~\cite{KNO}.
However this makes the trouble in UG, since then the cosmological constant must exist, in contrast to
the common understanding. To cancel the cosmological constant, the multiplier field $\la'$ should not be
a BRST daughter field. For this reason it was not possible to impose the physical state condition
even though the theory, at the Feynman graph level, seems to be well defined.

(2) Since UG has only the invariance under TDiff, we are allowed to gauge fix the invariance,
and the corresponding ghosts should satisfy the transverse condition off-shell. However in our previous formulation
the ghosts only satisfied it on-shell. We concluded that this is deeply connected with the problem (1).

Here in this paper we have just gauge fixed only the TDiff invariance, and there is no problem with
the subsidiary gauge-
conditions. Then what happens to the cosmological constant?
In GR in the unimodular gauge, it is impossible to absorb the cosmological constant $\Lambda$ into $\la'$
since $\la'$ is a BRST daughter field without VEV. We then have $\Lambda$ and $\la'$
separately in the action~\eqref{eq:UGaction1}, and $\Lambda$ gives the physical cosmological constant after imposing
the subsidiary condition~\cite{KNO}.
However in UG, we could consider the theory with cosmological constant but it has nothing to do with
the ``cosmological constant'' $\Lambda$ in the action.
Indeed, it is possible to absorb $\Lambda$ into $\la$ as in \eqref{eq:UGaction2}, and
the real cosmological constant appears as an integration constant in the field equation.
This is equivalent to specifying the VEV of the Lagrange multiplier field $\la$.
We should then define a new Lagrange multiplier field $\la''=\la-\lan \la\ran$ without VEV,
and it is this $\la''$ that falls into a daughter member of BRST quartet and must vanish by the subsidiary condition.
The earlier problem is resolved in this way.

In retrospect, though we have introduced antisymmetric tensor fields for the antighosts as well, this is not
anything that was required in the covariant quantization. It is true that the present formulation
possesses a symmetry between the ghosts and antighosts, but what is really necessary is to use them
only for the ghosts (not antighosts) to express the TDiff transformation parameter.
If we could gauge fix only the TDiff invariance without using such antisymmetric tensor fields for the antighosts,
there would not be new gauge invariance associated with the antighosts, and this would lead to simpler quantization
with fewer ghosts, though this leads to an asymmetric formulation in ghosts and antighosts.
In the accompanying paper~\cite{KNO2}, we will report the results in this direction.

\section*{Acknowledgment}

TK is supported in part by the JSPS KAKENHI Grant Number JP18K03659.
NO is supported in part by the Grant-in-Aid for Scientific Research Fund of the JSPS (C) Nos. 16K05331,
20K03980, and Taiwan MOST 110-2811-M-008-510.

\appendix

\section{Covariant Divergence of antisymmetric tensors}

Some elementary facts on covariant divergence of totally antisymmetric
tensors are given here.

Let $a^{\mu_1\mu_2\cdots\mu_n}$ generally denote rank-$n(\leq d)$ contravariant
tensors which are totally antisymmetric with respect to the $n$ indices
$\mu_1,\mu_2,\cdots,\mu_n$. Their covariant divergence have very simple expressions:
\begin{equation}
\nabla_\mu a^{\mu\nu_1\cdots\nu_n} =
\rtg^{-1}\partial_\mu(\rtg a^{\mu\nu_1\cdots\nu_n})\,.
\label{eq:A1}
\end{equation}
This is because the Christoffel connection has a particular form
\begin{equation}
\Gamma_{\mu\lambda}^\mu= \rtg^{-1}(\partial_\lambda\rtg),
\end{equation}
and so the covariant divergence of totally antisymmetric contravariant
tensors is calculated as
\begin{align}
\nabla_\mu a^{\mu\nu_1\cdots\nu_n}
&= \partial_\mu a^{\mu\nu_1\cdots\nu_n} + \Gamma_{\mu\lambda}^\mu a^{\lambda\nu_1\cdots\nu_n}
+\sum_{i=1}^n \Gamma_{\mu\lambda}^{\nu_i}a^{\mu\nu_1\cdots\overset{i\atop \vee}{\lambda}\cdots\nu_n} \nn
&=\partial_\mu a^{\mu\nu_1\cdots\nu_n} + \rtg^{-1}(\partial_\lambda\rtg) a^{\lambda\nu_1\cdots\nu_n}
= \rtg^{-1}\partial_\mu\left(\rtg a^{\mu\nu_1\cdots\nu_n}\right)\,.
\label{eq:A3}
\end{align}
Note that the last covariantization terms in the first line
vanish since the connection $\Gamma_{\mu\lambda}^{\nu}$ is $\mu$-$\lambda$ symmetric while
$a^{\mu\nu_1\cdots\lambda\cdots\nu_n}$ is $\mu$-$\lambda$ antisymmetric.

Next is the most useful property in our context:
\begin{equation}
\nabla_\mu\nabla_\nu a^{\mu\nu\rho_1\cdots\rho_n} = 0\,.
\label{eq:A4}
\end{equation}
For the contravariant tensor cases, this can most simply be proved
by using the above formula (\ref{eq:A1}); noting that
$\nabla_\nu a^{\mu\nu\rho_1\cdots\rho_n}$ is also a rank-$(n+1)$ totally
antisymmetric tensor,
\begin{align}
\nabla_\mu\nabla_\nu a^{\mu\nu\rho_1\cdots\rho_n}
&= \rtg^{-1}\partial_\mu\bigl( \rtg \nabla_\nu a^{\mu\nu\rho_1\cdots\rho_n} \bigr) \nn
&= \rtg^{-1}\partial_\mu\left( \rtg \bigl[
\rtg^{-1}\partial_\nu\left(\rtg a^{\mu\nu\rho_1\cdots\rho_n}\right)
\bigr] \right) \nn
&= \rtg^{-1}\partial_\mu\partial_\nu\left(\rtg a^{\mu\nu\rho_1\cdots\rho_n} \right) =0\,.
\end{align}

Actually the same form formula as this also holds for the covariant
antisymmetric tensors $A_{\mu_1\cdots\mu_n}$,
\begin{equation}
\nabla^\mu\nabla^\nu A_{\mu\nu\rho_1\cdots\rho_n}=0,
\end{equation}
although the simple formula like Eq.~(\ref{eq:A3}) does not exist.
This trivially follows since any covariant antisymmetric tensors can
be converted into contravariant tensors by multiplication of
metric tensors and metric tensors are freely commutative with covariant
derivatives.

The Hodge dual tensor $A_{\mu\nu_1\cdots\nu_n}$: \
Let us introduce a tensor $A_{\rho_1\cdots\rho_p}$ dual to
$a^{\mu\nu_1\cdots\nu_n}$ by
\begin{equation}
\rtg a^{\mu\nu_1\cdots\nu_n} = \pm (p!)^{-1}\varepsilon^{\mu\nu_1\cdots\nu_n\rho_1\cdots\rho_p}
A_{\rho_1\cdots\rho_p}\,,
\label{eq:A7}
\end{equation}
or, equivalently, by
\begin{equation}
A_{\rho_1\cdots\rho_p} = \mp \frac1{(n+1)!}
\rtg a^{\mu\nu_1\cdots\nu_n}
\varepsilon_{\mu\nu_1\cdots\nu_n\rho_1\cdots\rho_p} \ .
\end{equation}
Then, both sides of Eq.~(\ref{eq:A7}) are {\it tensor density} and
the simple divergence of them are again contravariant tensor density
by Eq.~(\ref{eq:A3}):
\begin{align}
\partial_\mu\rtg a^{\mu\nu_1\cdots\nu_n} &= \rtg \nabla_\mu a^{\mu\nu_1\cdots\nu_n} \nn
&=\pm (p!)^{-1}\varepsilon^{\mu\nu_1\cdots\nu_n\rho_1\cdots\rho_p}\partial_\mu A_{\rho_1\cdots\rho_p}
=\pm (p!)^{-1}\varepsilon^{\mu\nu_1\cdots\nu_n\rho_1\cdots\rho_p}\nabla_\mu A_{\rho_1\cdots\rho_p}\,.
\end{align}
The last equality should of course hold because of the covariant
divergence of (\ref{eq:A7}), and can also be proven directly as
\begin{equation}
\nabla_{\mu}A_{\rho_1\cdots\rho_p}= \partial_{\mu}A_{\rho_1\cdots\rho_p}-
\sum_{i=1}^p\Gamma^\lambda_{\mu\rho_i}A_{\rho_1\cdots\underset{\wedge\atop i}{\lambda}\cdots\rho_p}
\ \rightarrow\
\nabla_{[\mu}A_{\rho_1\cdots\rho_p]}= \partial_{[\mu}A_{\rho_1\cdots\rho_p]}\,.
\label{eq:covRotation}
\end{equation}
Indeed antisymmetrization with respect to the indices $\mu, \rho_1, \cdots, \rho_p$
eliminates the covariantization terms since $\Gamma^\lambda_{\mu\rho_i}$ is
$\mu$-$\rho_i$ symmetric.


\end{document}